\documentclass{aa}

\usepackage{amsmath}
\usepackage{mathtools}
\usepackage{graphicx}
\usepackage{natbib}
\usepackage{multirow}
\bibpunct{(}{)}{;}{a}{}{,}
\usepackage[varg]{txfonts}
\usepackage{mathrsfs}
\usepackage{xspace}
\usepackage{url}

\usepackage[usenames]{color}

\newcommand{\je}{IGR\,J18245$-$2452\xspace}

\newcommand{\swift}{\textsl{Swift}\xspace}
\newcommand{\integral}{\textsl{INTEGRAL}\xspace}
\newcommand{\xmm}{\textsl{XMM-Newton}\xspace}
\newcommand{\pn}{\textsl{EPIC-pn}\xspace}
\newcommand{\mos}{\textsl{EPIC-MOS}\xspace}
\newcommand{\rgs}{\textsl{RGS}\xspace}
\newcommand{\chandra}{\textsl{Chandra}\xspace}

\begin{document}

\title{``Hiccup'' accretion in the swinging pulsar \object{IGR\,J18245$-$2452}}

\author{C. Ferrigno
        \inst{1}
	\and 
        E. Bozzo\inst{1}
        \and
        A. Papitto\inst{2}
        \and
        N. Rea\inst{2,3}
        \and
        L. Pavan\inst{1}
        \and
        S. Campana\inst{4}
       \and
       M. Wieringa\inst{5}
        \and
        M. Filipovi\'c\inst{6}
        \and
       M. Falanga\inst{7,8}
       \and 
       L. Stella\inst{9}
      }

\authorrunning{C. Ferrigno et al.}
\titlerunning{``Hiccup'' accretion in the swinging pulsar IGR\,J18245$-$2452}
   \offprints{C. Ferrigno}

\institute{ISDC, Department of Astronomy,  Universit\'e de Gen\`eve, chemin d'\'Ecogia, 16, CH-1290 Versoix, Switzerland\\
	\email{Carlo.Ferrigno@unige.ch}
	\and
	Institute of Space Sciences (ICE; IEEC-CSIC), Campus UAB, Faculty of Science, Torre C5, Parell, 2a Planta, E- 08193 Barcelona, Spain
	\and
        Anton Pannekoek Institute, University of Amsterdam, Postbus 94249, NL-1090 GE Amsterdam, the Netherlands
        \and
        INAF-Osservatorio Astronomico di Brera, Via Bianchi 46, I-23807 Merate, Lecco, Italy
	\and
	CSIRO Astronomy and Space Science, Locked Bag 194, Narrabri NSW 2390, Australia
	\and
	University of Western Sydney, Locked Bag 1797, Penrith South DC, NSW 1797, Australia
        \and
        International Space Science Institute, Hallerstrasse 6, CH-3012 Bern, Switzerland
        \and
        International Space Science Institute in Beijing, No. 1, Nan Er Tiao, Zhong Guan Cun, Beijing, China
        \and
        INAF/Osservatorio Astronomico di Roma, via Frascati 33, I-00040 Monteporzio Catone, Roma, Italy
	 }

\date{Received ---; accepted ---}

\abstract{ {The source \je\ is the fifteenth discovered accreting millisecond
    X-ray pulsar and the first neutron star to show direct
    evidence of a transition between accretion- and rotation-powered
    emission states. These swings provided the strongest
    confirmation to date of the pulsar recycling scenario.}
  {During the two \xmm\ observations that were carried out while the
    source was in outburst in April 2013, \je\ displayed a unique and
    peculiar X-ray variability.} {In this
    work, we report on a detailed analysis of the \xmm\ data and focus
    on the timing and spectral variability of the
    source.  In the 0.4--11\,keV energy band, \je\ continuously switched
    between lower and higher intensity states, with typical variations
    in flux by factor of $\sim$100 on time scales as short as a few
    seconds. These variations in the source intensity were
    sometimes accompanied by dramatic spectral hardening, during
    which the X-ray power-law photon index varied from $\Gamma=$1.7 to
    $\Gamma=$0.9. The pulse profiles extracted at different
    count-rates, hardnesses, and energies also showed a complex variability. 
    These phenomena were never
    observed in accreting millisecond X-ray pulsars, at least not on such a short 
    time-scale.
    Fast variability was also found in the 5.5 and 9\,GHz ATCA radio observations 
    that were carried out
    for about 6\,h during the outburst.\\}  {
    We interpret the variability
        observed from \je in terms of a ``hiccup'' accretion phase,
    during which the accretion of material from the inner boundary of
    the Keplerian disk is reduced by the onset of
        centrifugal inhibition of accretion, possibly causing the
        launch of outflows. Changes across accretion and
    propeller regimes have been long predicted and
        reproduced by magnetohydrodynamic simulations of accreting millisecond X-ray
    pulsars, but have never observed to produce as extreme a variability as
    that shown by \je.}}
		

\maketitle

\section{Introduction}
The X-ray transient source 
\je was discovered in outburst by the hard X-ray imager
IBIS/ISGRI \citep{ubertini03, lebrun03} on-board \integral\ on 2013
March 28 during monitoring observations of the Galactic
Center \citep{eckert13}. The preliminary X-ray position provided by
ISGRI localized it within the globular cluster (GC) M~28.
This association was confirmed by follow-up observations carried out
with the \chandra\ ACIS-I telescope \citep{garmire03} and the
\swift\ XRT \citep{burrows05}.  A type I X-ray burst found in the XRT
data \citep{papitto2013a,linares13} firmly
established the source as an accreting neutron star X-ray binary
\citep[see also][]{serino13}.  Pulsations at 3.9\,ms were detected in
the X-ray flux recored from the source on April 3 and 13 by
\xmm,\; these made \je\  the fifteenth discovered
accreting millisecond X-ray pulsar (AMXP).  The delays in pulse-arrival 
times measured by \xmm\ led to the determination of the
source orbital period at 11.03\,hr \citep[][hereafter, Paper\,I]{papitto13}.  
The measured ephemeris of the system permitted us to
securely associate \je\ with a previously known radio millisecond
pulsar in M~28 (PSR\,J1824$-$2452I), thus making this source the first
millisecond pulsar to show direct evidence for swings between regimes with
emission powered by either rotation or accretion.  This result
provided the most secure confirmation to date of 
the so-called pulsar recycling scenario \citep{alpar82,radhakrishnan1982},
according to which old radio pulsars are secularly
spun-up to millisecond periods by mass accretion in low-mass
binary systems \citep[see][for the first evidence of a spin-up in AMXP]{falanga2005}. 
During the accretion phases, the material inflowing
from the companion star forms an accretion disk around the neutron star (NS) and
quenches the radio emission previously powered by the rotation of its magnetic field 
\citep[see][for previous indirect evidence of switches between these two 
phases in SAX~J1808.4$-$3659 and PSR\,J1023+0038, respectively]{burderi03,archibald09}.

\label{sec:intro} 
\begin{table}
\caption{Log of the \xmm\ observations.}
\begin{center}
\begin{tabular}{ccc}
\hline
\hline
OBSID & 0701981401 &  0701981501 \\
 &  \emph{(obs1)} &  \emph{(obs2)} \\
\hline
Start\tablefootmark{a} & 2013-04-03 23:31 & 2013-04-13 06:08\\
Stop\tablefootmark{a} & 2013-04-04 07:51 & 2013-04-14 01:43\\
\hline
\hline
\multicolumn{3}{c}{Effective exposure time (ks)}\\
\hline
\mos	 & 28.8 & 69.3\\
\pn	 & 26.8 & 67.3\\	  
\rgs	 & 29.0 & 69.5\\  
OM	 & 28.1& 67.1\\
\hline
\end{tabular}
\tablefoot{
\tablefoottext{a}{Times are in UTC.}}
\end{center}
\label{tab:obs}
\end{table}

In addition to these swings between the different emission regimes, \je\ 
displayed a peculiar variability in the X-ray
domain during its 2013 outburst, which was not observed before in other AMXPs. This variability
was particularly evident in the two \xmm\ target-of-opportunity observations 
carried out during the source outburst. In these observations, the X-ray
emission from \je\ repeatedly varied in
intensity by a factor of $\sim$100 on time-scales as short as a few
seconds. Here, we report on a detailed timing and spectroscopic
analysis of the \xmm\ data, aimed at investigating the nature and origin of the 
source X-ray variability.  

We describe the data analysis
technique in Sect.~\ref{sec:data} and report all the results in
Sect.~\ref{sec:results}. We present our discussions in
Sect.~\ref{sec:discussion} and propose in Sect.~\ref{sec:conclusions}
that the complex X-ray variability of \je\ arises in
an unstable (``hiccup'') accretion phase, during which the
inflowing material is able to penetrate the NS magnetosphere at
different latitudes and is often ejected instead of accreted onto the NS because of
the onset of the propeller mechanism \citep[][]{illarionov75,campana98}.

\begin{figure}
  \begin{center}
	\includegraphics[angle=0,width=0.5\textwidth]{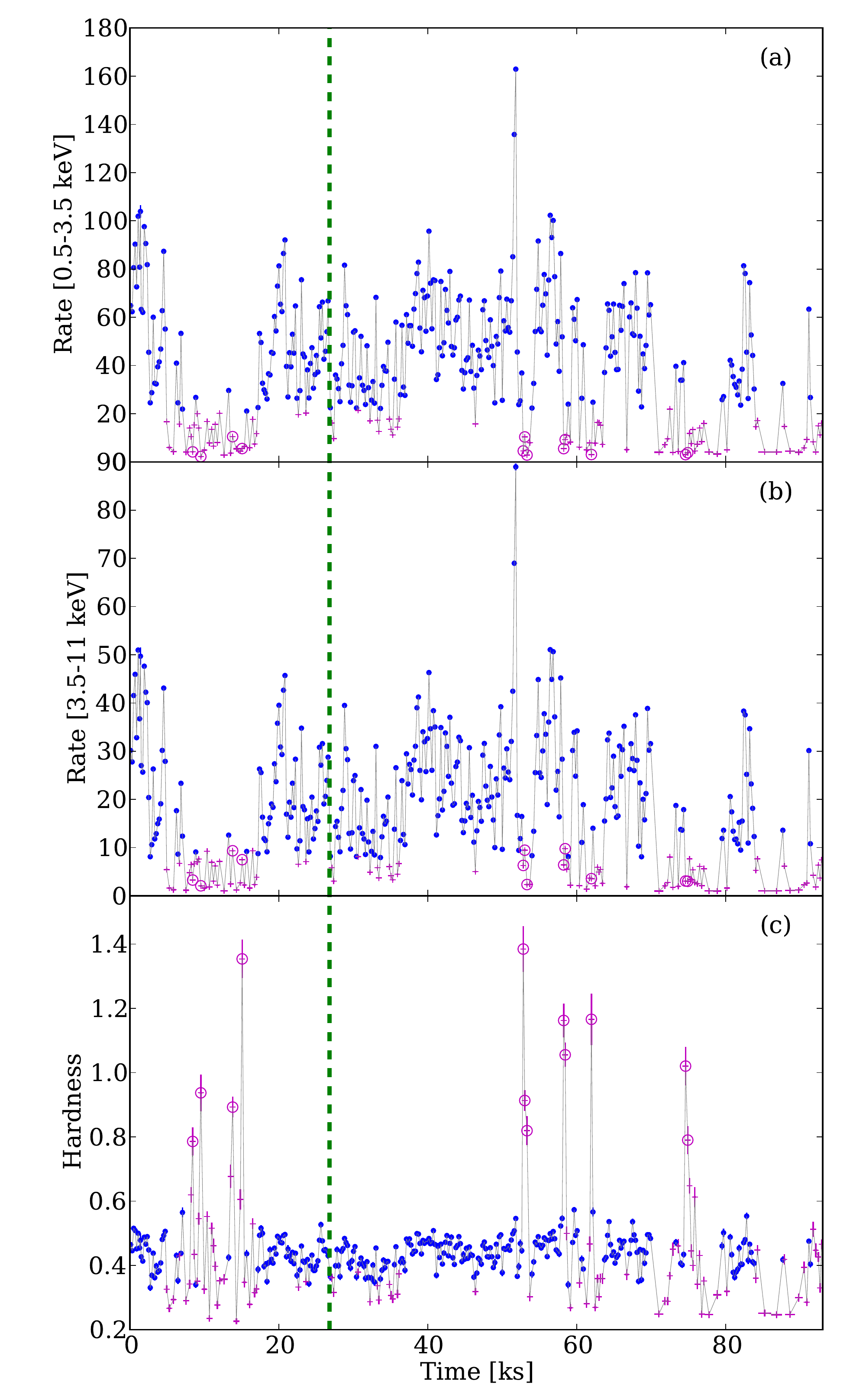}
	\caption{{\it Top panel}: \xmm lightcurve of \je\ extracted from obs1 and obs2 in the 0.5-3.5\,keV energy band. 
	The vertical dashed line represents the separation between 
	the two observations. Times of obs1 are measured starting from 00:25:22 on 2013 April 04 (TBD). For the second observation the start time is 
	2013 April 13 at 07:03:06 (TBD; note that times have been shifted by 26.3\,ks for plotting purposes). 
	{\it Middle panel}: same as before, but in the 3.5-11\,keV energy 
	range. This lightcurve was adaptively rebinned to achieve a signal-to-noise (S/N) ratio of 25 in each time bin, while 
	the minimum bins size was set to 200\,s (the same binning has been  
	used for the lightcurve in the soft energy band). {\it Bottom panel}: hardness ratio (HR) calculated as 
	the ratio of the hard and soft lightcurves.
	We plot with magenta symbols the time intervals in which the source count rate was lower than 30 cts/s, all the others are given in blue.
	The points with HR$>$0.7 are highlighted with an open circle.
	}
\label{fig:hr1}
\end{center}
\end{figure}

\section{Data analysis}
\label{sec:data}
The log of the \xmm\ observations we analyzed is provided
in Table~\ref{tab:obs}. In the following, we refer to observation
ID.\,0701981401 as obs1 and to observation ID.\,0701981501 as
obs2. In both cases, the \pn was operated in timing mode to attain
a timing resolution of 0.03\,ms and limit the effect of 
pile-up\footnote{See
  \url{http://xmm.esac.esa.int/external/xmm_user_support/
  documentation/uhb_2.2/node28.html.}}. The two \mos cameras were
operated in small-window mode and the \rgs in
standard mode (a thick filter was used for all the EPIC cameras
to screen the bright optical light from the globular cluster M~28).
  
Observation data files (ODFs) from obs1 and obs2 were processed to
produce calibrated event lists using the standard \xmm\ Science
Analysis System (v. 13.5.0). We used the {\sc epchan}, {\sc emproc},
and {\sc rgsproc} tasks to produce cleaned event files from the
\pn, the two \mos, and the \rgs instruments, respectively.  No high
flaring background time intervals were identified by following the SAS
science analysis threads\footnote{See
  \url{http://xmm.esac.esa.int/sas/current/documentation/threads}.}.
Therefore, we retained the entire available exposure time
for both obs1 and obs2.

In the \pn pipeline, we used the rate-dependent pulse height aptitude (PHA) correction to optimize 
the energy reconstruction of the events\footnote{See \url{http://xmm2.esac.esa.int/docs/documents/CAL-SRN-0299-1-1.ps.gz}.}.
The \pn light curves and spectra of the source were extracted by
using data from columns 26 to 47 (included) of the detector.  
The corresponding
backgrounds were extracted from columns 2 to 5 (included).  We
verified with the SAS task {\sc epaplot} that when the count rate from
the source exceeded $\sim$50 cts/s (0.5-11\,keV), the data extracted
from the two central columns of the \pn camera (37 and 38) were
affected by pile-up. We corrected for this effect by excluding data
from columns 37 and 38 during the spectral analysis. The response 
files for the \pn were generated taking into account the
pile-up problem according to the latest available SAS thread (see above);
the default point-spread function model was assumed.  \pn spectra were grouped
to have at least 40 counts in each energy bin 
and to oversample the energy resolution by
no more than a factor three.  The energy range is limited to 0.6--11\,keV, as at lower
energies the electronic noise distorts the spectrum\footnote{See
\url{http://xmm2.esac.esa.int/docs/documents/CAL-TN-0083.pdf}.}.
For the timing analysis, we obtained the photon
arrival times at the solar system barycenter using the {\sc barycen}
tool. Following the recommendations of the SAS user manual, we did not attempt to merge the spectra from the two observations 
characterized by a high count rate, but instead
fit them jointly\footnote{See \url{http://xmm.esac.esa.int/sas/current/documentation/threads/epic_merging.shtml}.}.

\begin{figure}
  \begin{center}
\resizebox{\hsize}{!}{
	\includegraphics[angle=0]{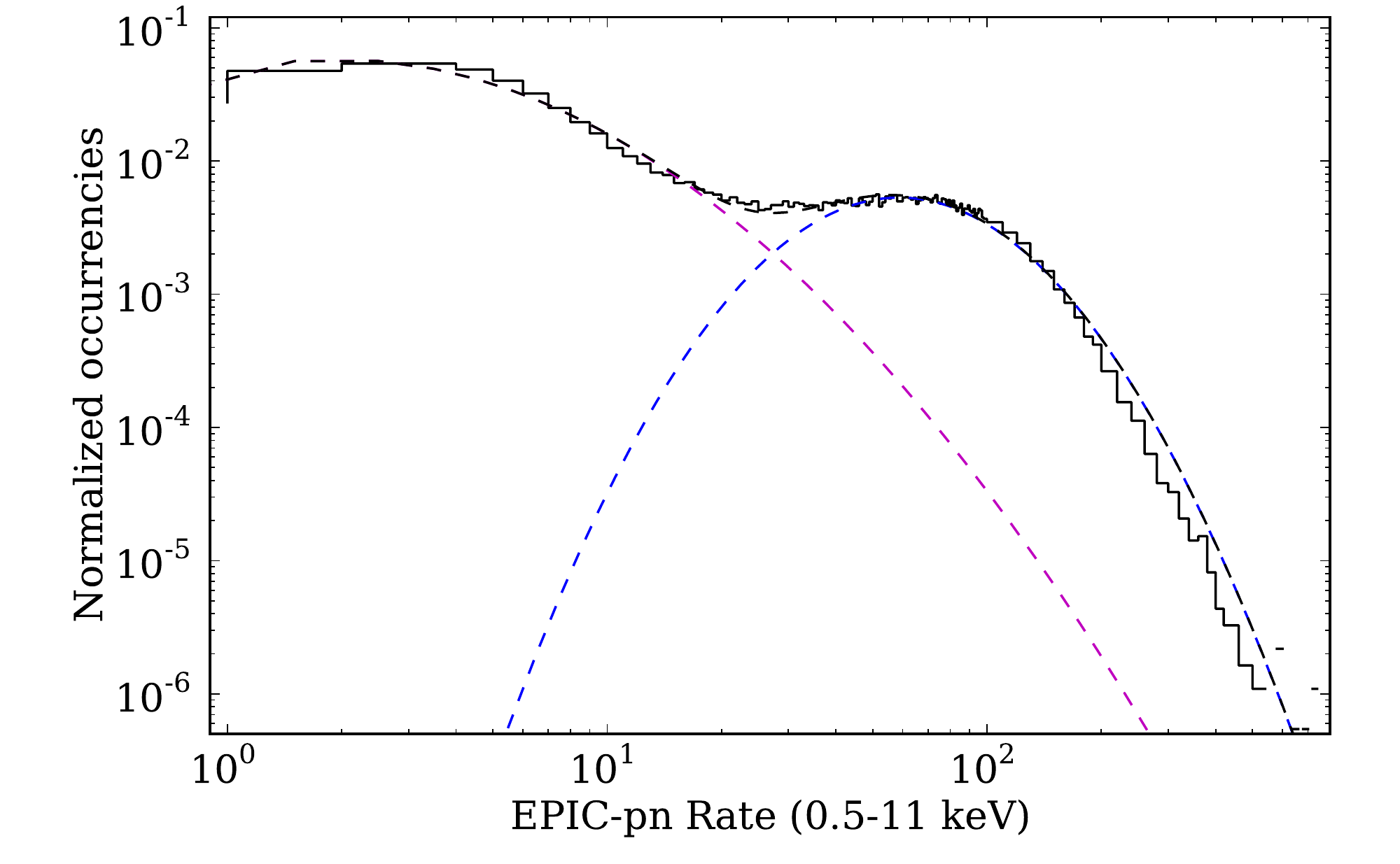}\hfill}
	\caption{Histogram of the count-rate derived from the combined lightcurves of the two observations 
	(energy range 0.5--11\,keV, time bin 1\,s). 
	The black solid line represents the total source count-rate.
	The black dashed line is the sum of two best-fit log-normal distributions, plotted in blue and magenta.}
\label{fig:histo}
\end{center}
\end{figure}

\begin{figure}
  \begin{center}
  \resizebox{\hsize}{!}{
	\includegraphics[angle=0]{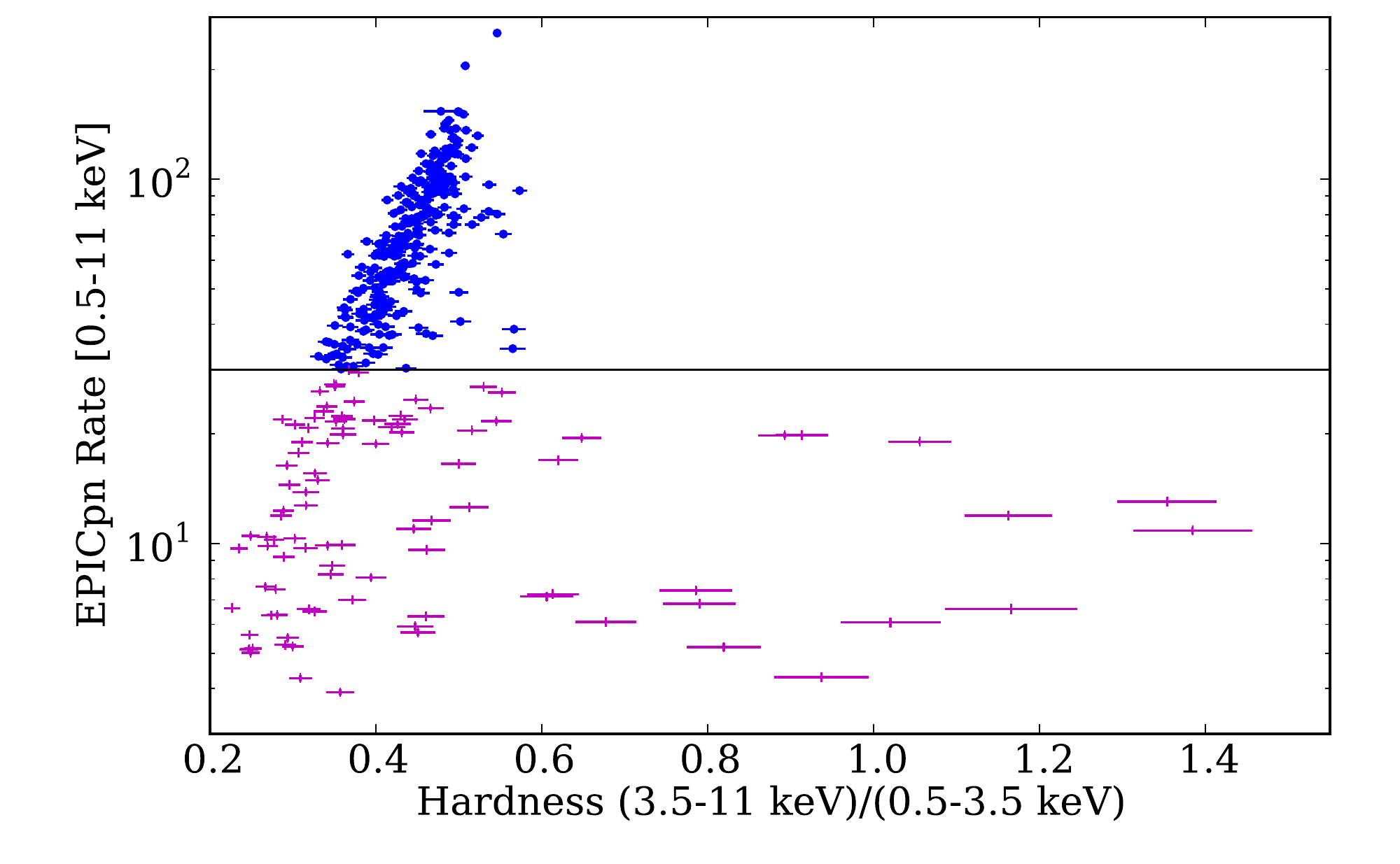}}
	\caption{Hardness-intensity diagram built by using all data in obs1 and obs2, which are displayed in Fig.~\ref{fig:hr1} (time bin of 200\,s). 
	The black solid line separates the different intensity states: the points represented in magenta and blue have a count rate lower and higher than 30 cts/s, respectively.}
\label{fig:hr-intensity}
\end{center}
\end{figure}


During our analysis, we found that data from the two MOS cameras were strongly affected by pile-up, and using these data did not improve 
any of the results presented in this paper (we verified {\it a posteriori} that all results extracted from the \mos data were compatible 
to within the large uncertainties with those obtained from the \pn). Therefore, we do not discuss the \mos data further. 
\begin{figure}
  \begin{center}
\resizebox{\hsize}{!}{
	\includegraphics[angle=0]{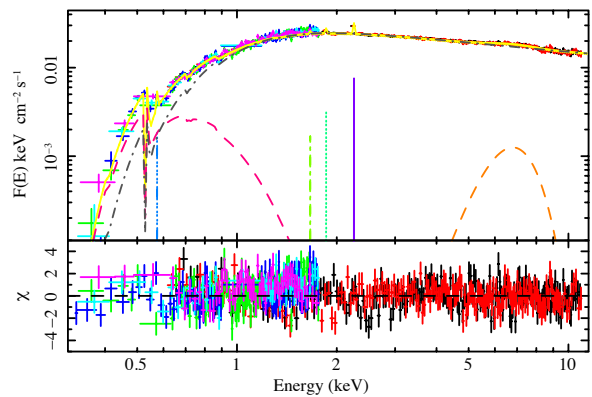}\hfill}
	\caption{Average unfolded spectra of \je in obs1 and obs2. Colors refer to the different observations 
	and instruments (red: \pn obs1; black: \pn obs2; green: \rgs1 obs2; blue: \rgs2 obs2; 
	light blue: \rgs1 obs1; magenta: \rgs2 obs1).
	The best-fit model is reported in Table\,\ref{tab:spec_all}. 
	We show in the lower panel the residuals from the best fit (see text for details).}
\label{fig:spec_all}
\end{center}
\end{figure}
\begin{table}
\caption{Results of the fits to the average spectra extracted by using all the available exposure time in obs1 and obs2. The model used for the fits 
comprises a photoelectric absorption with variable abundances of iron and oxygen \citep[\texttt{tbnew\_feo},][]{wilms01}, a thermal Comptonization 
(\texttt{compTH} in {\sc xspec}), a thermal disk black-body, a narrow Gaussian line 
representing the blend pf \ion{O}{VII} emission lines, and a broad Gaussian iron line for the iron emission.
The spectra of the two observations were fit together to achieve the most stringent constraints 
on the X-ray emission properties.}
\begin{center}
\begin{tabular}{cr@{}lc}
\hline
\hline
Parameter &  \multicolumn{2}{c}{Value} & Unit \\
\hline
\smallskip
$N_\mathrm{H}$ & 43.1 & $^{+0.5}_{-1.7}$  & $\times 10^{20} \mathrm{cm}^{-2}$\\
\smallskip
$A_\mathrm{O}$ & 0.96 & $\pm$0.01 & Solar\\
\smallskip
$A_\mathrm{Fe}$ & 0.75 & $\pm$0.05 & Solar\\
\smallskip
$kT_\mathrm{in}$\tablefootmark{a} & 0.127 & $\pm$0.003 & keV \\
\smallskip
$r_\mathrm{in}\sqrt{\cos i}$\tablefootmark{a} & 66.4 & $^{+11.1}_{-0.7}$ & km \\
\smallskip
$kT_\mathrm{nthComp}$\tablefootmark{b} & 0.257 & $\pm$0.010 & keV \\
\smallskip
$\Gamma$ & 1.428 & $\pm$0.003 & \\
\smallskip
$N_\mathrm{nthComp}$ & 3.035 &$\pm$0.005 & $\times 10^{-2}$\\
\smallskip
$E_\mathrm{Fe}$ & 6.6 & $\pm$0.1 & keV \\
\smallskip
$\sigma_\mathrm{Fe}$ & 1.08 & $\pm0.11$ & keV \\
\smallskip
$N_\mathrm{Fe}$ & 5.1 & $\pm0.4 $ & $\times10^{-4}\,\mathrm{cm^{-2}s^{-1}}$ \\
\smallskip
$E_\mathrm{\ion{O}{VII}}$ & 0.5719 & $^{+0.0004}_{-0.0010}$ & keV \\
\smallskip
$\sigma_\mathrm{\ion{O}{VII}}$ & 0 & -- & keV \\
\smallskip
$N_\mathrm{\ion{O}{VII}}$ & 37 & $\pm11 $ & $\times10^{-4}\,\mathrm{cm^{-2}s^{-1}}$ \\
\smallskip
Flux1 [0.5--11\,keV]\tablefootmark{c} & 3.123 & $\pm$0.007 &$\times10^{-10} \,\mathrm{erg\,cm^{-2}s^{-1}}$ \\
\smallskip
Flux2 [0.5--11\,keV]\tablefootmark{c} & 3.033 & $\pm$0.010 &$\times10^{-10} \,\mathrm{erg\,cm^{-2}s^{-1}}$ \\
\smallskip
$\chi^2_\mathrm{red}$/d.o.f & 1.1297 & /4067 & \\
\hline
\smallskip
$E_1$\tablefootmark{d}  & 1.67&$^{+  0.04}_{-  0.05}$ & keV \\
\smallskip
$N_1$\tablefootmark{d} & 2.8 & $\pm$1.0& $\times10^{-5} \,\mathrm{cm^{-2}s^{-1}}$ \\
\smallskip
$E_2$\tablefootmark{d} & 1.85&$\pm0.03$ & keV\\
\smallskip
$N_2$\tablefootmark{d} & 5 & $\pm$ 2& $\times10^{-5} \,\mathrm{cm^{-2}s^{-1}}$ \\
\smallskip
$E_3$\tablefootmark{d} &  2.250  &$\pm$0.010 & keV \\
\smallskip
$N_3$\tablefootmark{d} & 11.3 & $\pm$1.6 & $\times10^{-5} \,\mathrm{cm^{-2}s^{-1}}$ \\
\hline
\end{tabular}
\tablefoot{The spectra are coupled with cross-calibration constants, whose values are between 0.971$\pm$0.004 and 1.09$\pm$0.01.\\
\tablefoottext{a}{Temperature and apparent radius of the disk black-body component.}\\
\tablefoottext{b}{Seed photon temperature of the nthComp model.}\\
\tablefoottext{c}{The flux is computed from the \pn spectra and the uncertainties are only statistical, the absolute flux calibration of an X-ray camera has typical uncertainty of a few percent \citep[e.g.,][]{romano2005}.}\\
\tablefoottext{d}{Lines due to calibration problems.}}
\end{center}
\label{tab:spec_all}
\end{table}

The \rgs spectra were extracted using standard
techniques in the energy range 0.4--1.8\,keV, where the 
calibration of the effective area achieves a 3\% accuracy \citep{kaastra2009}\footnote{See
  \url{http://xmm.esac.esa.int/sas/current/documentation/threads/rgs_thread.shtml}.}.  The \rgs spectra were rebinned to have 20
counts in each energy bin and obtain a reliable measurement of the
absorption column density in the direction of the source. For the hardness-resolved spectra,
extracted when the source was dimmer (Sect.~\ref{sec:variability}), we combined the two RGS units to increase the signal to noise ration (S/N).
The analysis of the \ion{O}{VII} triplet, which is detected as a single line in these rebinned \rgs
spectra, is discussed in Appendix\,\ref{sec:appendix}.
Unless otherwise stated, spectral fitting was
carried out by using Xspec v12.8 and uncertainties are given 
throughout our paper at 90\%~c.l..

\section{Results}
\label{sec:results}

The lightcurves of the two \xmm observations are shown in
Fig.\,\ref{fig:hr1}. In both observations, 
the source displays a very prominent
variability (all lightcurves are
background subtracted and corrected for instrumental effects,
e.g., vignetting).  During several relatively extended
periods of time (a few thousand seconds) 
the average source intensity was lower than in the rest of the observation,
retaining a remarkable variability.  A
single low-intensity period in obs1 began $\sim$5\,ks after the start of
the observation and lasted for about 13\,ks.  Multiple low-intensity 
periods occurred in obs2.  We show in panel (c) of Fig.\,\ref{fig:hr1} the
hardness ratio (HR) of the lightcurves extracted in the soft
(0.5--3.5\,keV) and hard (3.5--11\,keV) energy bands 
\citep[see][for details about the HR computation]{bozzo11,bozzo13}.
The HR displayed only a moderate variability during most of the
observations (between $\sim$0.3 and 0.5), but underwent rapid and strong variations during the low
intensity periods. We folded the lightcurves along 
the 2.3 orbits covered by \xmm, but no obvious orbital dependency 
of these phenomena was found.
 
To characterize the distribution of states at the highest
achievable timing resolution without facing shot-noise limitations, we
built the histogram of the source count rate accumulated in
time bins of one second.  By inspecting these lightcurves,
not shown here for brevity, we verified that the \pn count-rate showed
switches from $\sim$3--5\,cts/s to $\gtrsim$500 cts/s on time
scales as short as a few seconds (0.5-11\,keV).
The count-rate histogram is characterized by two broad peaks
that can be described by two log-normal distributions with centroids 
$5.48\pm0.06$ and $79.5\pm0.3$\,cts/s, and widths $2.75\pm0.04$ and
$1.74\pm0.03$\,cts/s, respectively ($\chi^2_\mathrm{red}=6.6$ for 121
d.o.f.). Because of this bi-modal
distribution, we
draw the separation between dimmer and brighter states at 30 cts/s, where the two
distributions cross. 
We represent in Fig.~\ref{fig:histo} 
the dimmer state in \emph{magenta} and marked the others in \emph{blue}; 
the same color coding is maintained for the rest of the paper and is used to identify the two states.
As can be seen in Fig.\,\ref{fig:hr1}, this division tracks
the intervals of lower activity relatively well, which are
characterized by remarkable swings of the hardness ratio and the periods of
more intense activity at nearly constant hardness. 

Fig.\,\ref{fig:hr-intensity} shows the hardness-intensity
diagram (HID) of the source realized by using the same color coding as 
of Figs.~\ref{fig:hr1} and \ref{fig:histo}.
We note that the blue points form a first branch that is characterized
by increasing hardness (HR) as a function
of the source intensity and saturates at HR$\sim$0.5--0.6; 
the source spends about 60\% of the time in this branch.
The magenta points are placed on a second branch that is characterized
by an average lower intensity and reaches
hardness values of 1.4. The source spends about 40\% of the time in this state. 

It is important to remark that the picture of the source variability emerging from Fig.~\ref{fig:hr1}
is somehow smoothed by the relatively large time bins. When the source light curve 
is extracted with a time bin of 1\,s, the swings are much more pronounced 
and reach count rates as high as 700 cts/s (Fig.~\ref{fig:histo}). We verified a posteriori 
that the time intervals corresponding to the magenta and blue states identified using the time bins of 200\,s 
track the different variability pattern of the source reasonably well. 
This is also valid for other similar choices of binning.

To understand the nature of the HR variations shown by the light curve analysis,
we first extracted the average \pn and \rgs spectra (Sect.\,\ref{sec:spectral}).
Then, we performed a rate- and HR-resolved spectral analysis for the blue and magenta states separately, 
using Fig. \ref{fig:hr-intensity} as guidance
(see Sect.\,\ref{sec:variability} and \ref{sec:pulse_profile}).
\begin{table*}
\caption{Spectral parameters of the best-fit models to the rate-resolved spectra of \je during the blue states.}
\begin{center}
\begin{tabular}{cr@{}lr@{}lr@{}lr@{}lc}
\hline
\hline
Rate\tablefootmark{a} &  \multicolumn{2}{c}{30--50} & \multicolumn{2}{c}{50--75} & \multicolumn{2}{c}{75--100} & \multicolumn{2}{c}{$>$100} & cts/s \\
\hline
\smallskip
$N_\mathrm{H}$ 	&    45 & $\pm8$   &   40 & $\pm 6$ &    46 & $\pm 7$  &   41 & $\pm6$	 &  $\times 10^{20} \mathrm{cm}^{-2}$\\
\smallskip
$kT_\mathrm{in}$ &  0.119 & $^{+  0.017}_{-  0.013}$	&    0.127 & $^{+  0.014}_{-  0.012}$	&    0.127 & $^{+  0.015}_{-  0.016}$	&    0.142 & $^{+  0.016}_{-  0.013}$ & keV \\
\smallskip 
$r_\mathrm{in}\sqrt{\cos i}$ &  85 & $^{+98}_{-31}$ & 69 & $^{+62}_{-22}$ & 96 & $^{+82}_{-31}$ & 70 & $^{+51}_{-22}$ & km \\
\smallskip
$E_\mathrm{Fe}$ 	&    6.9 & $\pm0.3$   &   6.4 & $\pm0.2$ &    6.6 & $\pm0.3$  &    6.6 & $\pm0.2$	 &  keV \\
\smallskip
$\sigma_\mathrm{Fe}$ 	&        1.0&-                            &           1.0&-                      &    1.3 &$^{+  0.8}_{-0.5}$  &    0.9&$^{+ 0.5}_{-0.2}$	 &  keV \\
\smallskip
$N_\mathrm{Fe}$ 	&    8&$\pm2$   &   10 & $\pm2$ &    10 & $^{+13}_{-4}$  &    11 & $^{+7}_{-3}$	 &  $\times10^{-4}\,\mathrm{cm^{-2}s^{-1}}$ \\
\smallskip
$\Gamma$ 	&  1.55& $\pm0.02$   &  1.472 & $\pm0.013$  &    1.42&$\pm0.02$  &    1.375&$^{+  0.016}_{-  0.011}$	 &  \\
\smallskip
$kT_\mathrm{nthComp}$ &  0.24 & $\pm0.04$   &  0.27 & $\pm0.03$  &    0.26 & $^{+  0.04}_{-  0.06}$  &    0.31 & $^{+0.1}_{-0.3}$	 & keV \\
\smallskip
$N_\mathrm{nthComp}$ 	&   2.7 &$\pm0.3$   &  3.9 &$\pm0.4$ &    5.4 & $\pm0.6$   &  6.6 &$\pm0.5$	 & $\times 10^{-2}$\\
\smallskip
Exp1\tablefootmark{b}  &  3.2 & &  5.3   & &  3.8    & & 2.8  & & ks\\
\smallskip
Exp2\tablefootmark{b} &  8.3 & &  11.6 & &  11.0 & &  7.3   & & ks\\
\smallskip
Flux\tablefootmark{c} &  2.23 & $\pm$0.01 & 3.49 & $\pm$0.02 & 5.13 & $\pm$0.03 & 7.86 & $\pm$0.05 &$\times10^{-10} \,\mathrm{erg\,cm^{-2}s^{-1}}$ \\
\smallskip
$\chi^2_\mathrm{red}$/d.o.f.\tablefootmark{d} & 1.051 &/1090 & 1.012 &/1569 & 0.940 &/1722 & 1.048 &/1782 & \\
\hline
\smallskip
$E_3$\tablefootmark{e} 	&    2.21 & $\pm0.05$   &  2.26 & $\pm0.02$  &    2.24 & $\pm0.02$  &    2.258 & $\pm0.015$	 &  keV \\
\smallskip
$N_3$\tablefootmark{e} 	&    5.1 & $\pm3.4$   &  1.3 & $\pm0.5$  &    2.0 & $\pm0.7$  &    4.1 & $\pm0.9$	 &  $\times10^{-4} \,\mathrm{cm^{-2}s^{-1}}$ \\
\hline
\end{tabular}
\tablefoot{
\tablefoottext{a}{\pn count rate in the 0.5--11\,keV energy range. The time selection is operated on the \pn lightcurves binned at 200\,s.}\\
\tablefoottext{b}{Exp1 and Exp2 are the \pn exposures during obs1 and obs2.}\\
\tablefoottext{c}{The absorbed flux is computed from the \pn spectrum of obs2 in the 0.5--10\,keV range (uncertainties are only statistical).}\\
\tablefoottext{d}{the intercalibration constants between the instruments varied within a limited range ($\pm10$\%).}\\
\tablefoottext{e}{Feature introduced to account for calibration problems.}}
\end{center}
\label{tab:spec_rate}
\end{table*}

\subsection{Average spectrum}
\label{sec:spectral}

The combined \rgs+\pn spectrum (0.4--1.8\,keV and 0.6--11\,keV) 
extracted by using the total exposure time
available in obs1 and obs2 is shown in Fig.\,\ref{fig:spec_all}.  For
both observations, we adopted 
a spectral model that comprises a photoelectric absorption with variable
abundances of iron and oxygen \citep[{\sc tbnew\_feo};][]{wilms01}, a thermal
Comptonization ({\sc compTH} in {\sc xspec}, the electron temperature was
fixed to 50\,keV), and three components due to the accretion disk: 
a broad Gaussian line, a disk black-body, and a narrow 
line corresponding to the non-resolved \ion{O}{VII} helium-like triplet (see Appendix A). 
Because of \pn calibration problems related to the energy scale in timing mode (see
also Sect.\,\ref{sec:data}), we introduced in the fit three
Gaussian lines with zero width and centroid energies of 1.5, 1.8, and
2.2\,keV. These energies correspond to pronounced edges in the instrument effective area\footnote{See
\url{http://xmm2.esac.esa.int/docs/documents/CAL-TN-0018.pdf}.}.  
We verified that
the relatively large $\chi^2_\mathrm{red}=1.13$ for 4067 d.o.f.
is mostly due to residual calibration uncertainties.

The results obtained by separately fitting the spectra of obs1
and obs2 with the model above are similar to within the uncertainties. 
We therefore report in Table\,\ref{tab:spec_all} only the
results obtained by simultaneously fitting all spectra with the same
model. 

\subsection{Spectral variability}
\label{sec:variability}

We first investigate the spectral variability of the blue states.
To this aim, we extracted four source spectra at different intervals of \pn
count rates higher than 30\,cts/s,
as reported in Table~\ref{tab:spec_rate}. Simultaneous \rgs1 and \rgs2
spectra were also extracted and fit together with the \pn spectra 
using the same model adopted in
Sect.\,\ref{sec:spectral}.  
With the reduced statistics of these spectra, only the edge 
at 2.2\,keV produced an observable feature in the spectrum, modeled as a 
zero-width Gaussian line.
We introduced in the fits normalization constants to account for
calibration uncertainties between the instruments and slight differences in the distribution
of source fluxes during the selected time intervals.
We checked that all parameters remained
compatible to within the uncertainties in the two observations, if
left free to vary in the fits. A broad Gaussian line at $\sim$6.4\,keV
was introduced, similarly as for the average spectrum.  
The results reported in Table~\ref{tab:spec_rate} show
a significant spectral variability within the blue states,
which appears to be caused by a harder asymptotic power-law photon index 
of the Comptonized spectrum for increasing count rates.

\begin{table*}
\caption{Reduced $\chi^2$ and degrees of freedom of various models for hardness-resolved spectra. }
\begin{center}
\begin{tabular}{cr@{}lr@{}lr@{}lr@{}lr@{}ll}
\hline
\hline
Model/HR\tablefootmark{a} & \multicolumn{2}{c}{$<$0.25}  & \multicolumn{2}{c}{0.25-0.33}  & \multicolumn{2}{c}{0.33-0.43}  & \multicolumn{2}{c}{0.43--0.70} &  \multicolumn{2}{c}{$>$0.70} &  \\
\hline
nthcomp & 1.192&/388 & 1.312&/496 & 1.178&/591 & 1.029&/392 & 2.063&/335 &  \\
nthcomp+BB & 1.053&/386 & 1.100&/494 & 1.034&/589 & 1.042&/390 & 1.161&/333 &  \\
nthcomp+diskBB & 1.040&/386 & 1.095&/494 & 1.179&/589 & 1.022&/390 & 1.161&/333 &  \\
nthcomp+PL & 1.093&/386 & 1.096&/494 & 1.107&/589 & 1.020&/390 & 1.183&/333 &  \\
Partial covering PL & 1.111&/387 & 1.381&/495 & 1.149&/590 & 1.040&/391 & 1.167&/334 &  \\

\hline
Exp$_\mathrm{1}$\tablefootmark{b} & \multicolumn{2}{c}{2.75} & \multicolumn{2}{c}{1.58} & \multicolumn{2}{c}{3.94} & \multicolumn{2}{c}{1.57} & \multicolumn{2}{c}{1.77} & ks \\
Exp$_\mathrm{2}$\tablefootmark{b} & \multicolumn{2}{c}{7.28} & \multicolumn{2}{c}{7.48} & \multicolumn{2}{c}{4.92} & \multicolumn{2}{c}{3.78} & \multicolumn{2}{c}{2.75} & ks \\
\hline
\end{tabular}
\label{tab:chi}
\tablefoot{
\tablefoottext{a}{Hardness ratio (HR) computed as in Fig.\,\ref{fig:hr1} for count rates lower than 30 cts/s.}
\tablefoottext{b}{Exposures of the \pn camera in the two observations. The \rgs exposures are similar, but not strictly 
equal because of different screening criteria and different operational window in obs1 and obs2.}}
\end{center}
\label{tab:chi_HR}
\end{table*}

\begin{figure}
  \begin{center}
	\includegraphics[angle=0,width=0.5\textwidth]{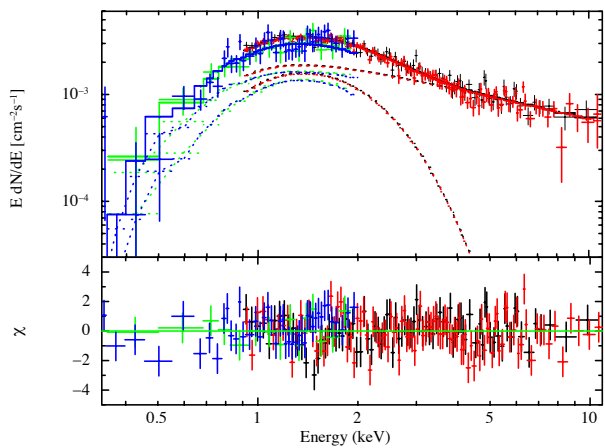}
	\caption{\je\ unfolded spectrum of the soft magenta state (HR$<$0.25). 
	The best-fit model (see Table\,\ref{tab:spec_hardness}) and the residuals from this fit 
	are shown as well. Colors refer to the different observations and instruments (black: \pn obs1; 
	red: \pn obs2; green: \rgs obs1; blue: \rgs obs2).}
\label{fig:spec_lowHR}
\end{center}
\end{figure}

\begin{figure}
  \begin{center}
	\includegraphics[angle=0,width=0.5\textwidth]{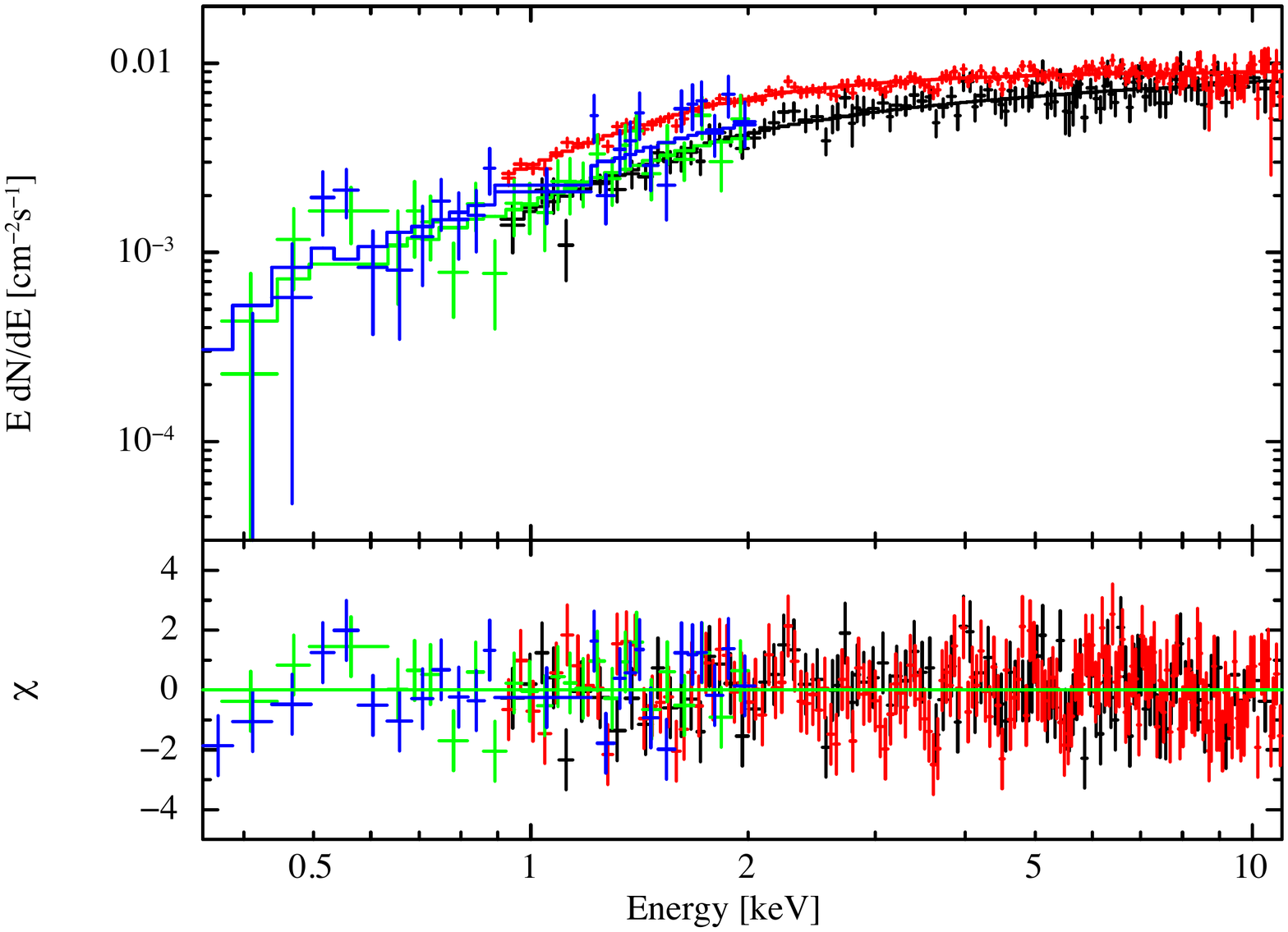}
	\caption{\je\ unfolded spectrum of the hard magenta state (HR$>$0.7). The best-fit model (see Table\,\ref{tab:spec_hardness}) and the residuals from this fit are shown as well.
	Colors refer to the different observations and instruments (black: \pn obs1; red: \pn obs2; green: \rgs obs1; blue: \rgs obs2).}
\label{fig:spec_highHR}
\end{center}
\end{figure}

Because the magenta states span a relatively small range of source count rates ($<$30 cts/s) 
while the source HR displays remarkable 
variability, we investigated spectral changes in these states by carrying out 
a HR-resolved (instead of count-rate-resolved) spectral 
analysis. From the results in Fig.\,\ref{fig:hr1}, we extracted the source \rgs and \pn 
spectra of the magenta states in 
obs1 and obs2 in five HR intervals. 
In almost all cases, using a simple absorbed {\sc nthComp} model resulted 
in a very poor fit with evident residuals emerging 
along the entire 0.4--11\,keV energy band. We therefore tried to improve 
the fits by adding a black-body, a disk black-body, or
a power-law to the {\sc nthComp} component. 
We also used an alternative model with a power-law and a partially covering absorber.  
Most of these models gave statistically equivalent fits to the HR-resolved 
spectra (see Table\,\ref{tab:chi_HR}). The {\sc nthComp}+power-law model 
was considered less suitable because in this case the power-law 
photon index displayed unphysically large swings in the different HR-resolved spectra. 
A model comprising the {\sc nthComp} component and a black-body 
was found to provide a more reliable description of all spectra with $HR<0.7$, 
because minor adjustments in the temperature and radius of the black-body 
were required to obtain $\chi_{\rm red}^2$$\simeq$1. 
An example is shown in Fig.~\ref{fig:spec_lowHR}. 
The radius and temperature (2.5\,km and 0.4\,keV) of the black-body component are reminiscent of those 
typically observed from hot spots on neutron-star surfaces. We verified that adding of a colder and larger thermal component, similar 
to that measured during the blue state and associated with the accretion disk around the neutron star, did not significantly improve any 
of the HR-resolved spectra extracted during the magenta states. The only exception was the spectrum extracted at HR$>$0.7. 
In this case, the parameters of the 
thermal component obtained from the fit were compatible with those expected from the inner boundary of an accretion disk, but  
the temperature of seed photons in the {\sc nthComp} component was far too high (a factor $>$2) 
and difficult to be reconciled 
with the nearly constant values obtained from the other HR-resolved spectra. 
We found that a partially covering model instead provided a reasonable description of this very hard and peculiar spectrum (Fig.~\ref{fig:spec_highHR}).
All the results obtained from the fits to the HR-resolved spectra are summarized in Table\,\ref{tab:spec_hardness} and are discussed 
in Sect.\,\ref{sec:discussion}.

We verified that in none of the count-rate and HR-resolved spectra extracted from the magenta states adding 
a broad Gaussian iron line significantly improved the fits (most likely because of the relatively limited statistics of these data).
The upper limits obtained on the normalization of the line were in 
all cases largely compatible with the values reported in Table\,\ref{tab:spec_all}. We also checked {\it a posteriori} that the spectra 
extracted at the same count-rate or HR intervals in obs1 and obs2 would give results compatible (despite the larger uncertainties) 
with those reported in Tables\,\ref{tab:spec_rate} and \ref{tab:spec_hardness} if fit independently in {\sc Xspec}.

From the results of the spectral fits we obtained an approximate conversion 
of the source count-rate into flux and determined the average flux of the blue and magenta states 
to be  $3\times10^{-11}\,\mathrm{erg\,s^{-1}\,cm^{-2}}$ and $4.0\times10^{-10}\,\mathrm{erg\,s^{-1}\,cm^{-2}}$.  
These correspond to a luminosity of $1\times10^{35}\,\mathrm{erg\,s^{-1}}$ and $1.4\times10^{36}\,\mathrm{erg\,s^{-1}}$ at 5.5\,kpc.

\begin{table*}
\caption{Parameters of the best-fit models to the HR-resolved spectra of the magenta using the 
{\sc nthComp} plus black-body. We also show the results obtained by adopting a partially covering power-law model 
for the spectrum extracted at HR$>0.7$.}
\begin{center}
\begin{tabular}{cr@{}lr@{}lr@{}lr@{}lr@{}lr@{}ll}
\hline
\hline
\smallskip
Hardness\tablefootmark{a} & \multicolumn{2}{c}{$<0.25$} & \multicolumn{2}{c}{0.25-0.33} & \multicolumn{2}{c}{0.33-0.43} & \multicolumn{2}{c}{0.43-0.70}   & \multicolumn{2}{c}{$>0.70$\tablefootmark{b}} & \multicolumn{2}{c}{$>0.70$\tablefootmark{c}} &  \\
\hline
\smallskip
$N_\mathrm{H}$      &    23 & $\pm7$   &    20 & $\pm7$   &    16 & $\pm4$   &   27 & $^{+3}_{-2}$   &    $8$ &$^{+8}_{-6}$ &   11 & $\pm8$  & $\times10^{20} \mathrm{cm}^{-2}$  \\
\smallskip
$kT_\mathrm{BB}$      &    0.40 & $\pm0.02$   &    0.46 & $^{+  0.02}_{-  0.04}$   &    0.43 & $^{+  0.06}_{-  0.04}$   &    0.75 & $^{+  0.07}_{-  0.05}$   &    0.010 & $^{+0.05}_{-0.04}$\tablefootmark{b} & -&- & keV \\
\smallskip
$R_\mathrm{BB}$    &   2.5 & $\pm0.3$   &   2.4 & $^{+  0.3}_{-  0.2}$   &   2.8 & $^{+ 1.8}_{- 0.6}$   &    0.86 & $^{+0.15}_{- 0.17}$   &  23 & $^{+205}_{-10}$\tablefootmark{b} & -&- & km\\
\smallskip
$N_\mathrm{H,pc}$ & -&- & -&- & -&- & -&- & -&- & 1.6 & $^{+0.5}_{-0.3}$ & $\times 10^{22} \mathrm{cm}^{-2}$\\
\smallskip
$F_\mathrm{pc}$ & -&- & -&- & -&- & -&- & -&- & 0.64 & $\pm0.08$ & \\
\smallskip
$\Gamma$  &    1.71 & $\pm0.12$   &    1.40 & $\pm0.05$   &    1.35 & $\pm0.04$   &    1.12 & $^{+  0.05}_{-  0.03}$   &    1.13  & $^{+0.07}_{-0.05}$ & 0.90 & $\pm0.04$ &  \\
\smallskip
$kT_\textrm{nthComp}$   &    0.19 & $^{+  0.11}_{-  0.08}$   &    0.22 & $\pm0.14$   &    0.4 & $^{+  0.4}_{-  0.1}$   &    $<0.1$ &   &    0.90 & $\pm0.08$ & -&- & keV \\
\smallskip
$N_\textrm{nthComp,PL}$\tablefootmark{d}       &    2.7 & $\pm0.8$   &    5.8 & $^{+  1.3}_{-  1.4} $   &    6.8 & $^{+  1.6}_{-  3.9} $   &   7.1 & $^{+  0.8}_{-  0.7}$   &    1.40 & $^{+0.12}_{-0.09} $ &  4.0 & $^{+  0.4}_{-  0.3} $ & $\times10^{-3}$ \\
\smallskip
$C2_\mathrm{P}$\tablefootmark{e}  &    0.98 & $\pm0.02$ &  0.949 & $\pm0.015$ & 0.902 & $\pm0.009$ &       0.64 & $\pm0.01$ & 1.26 & $\pm0.03$ & 1.26 & $\pm0.03$ & \\
\smallskip
$C1_\mathrm{R}$\tablefootmark{e} &   0.87 & $\pm0.07$ & 0.94 & $\pm0.07$ & 0.858 & $\pm0.035$ & 0.84 & $\pm0.07$ & 0.88 & $\pm0.12$ &  0.90 & $\pm0.13$ &    \\
\smallskip
$C2_\mathrm{R}$\tablefootmark{e} &   0.83 & $\pm0.04$ & 0.81 & $\pm0.03$ & 0.81 & $\pm0.03$ & 0.58 & $\pm0.04$ & 1.05 & $\pm0.12$ &    1.07 & $\pm0.12$ &  \\
\smallskip
Flux\tablefootmark{f} &  2.06 &$\pm0.04$ & 6.26 &$\pm0.07$ & 9.75 &$\pm0.10$ &  7.10 &$\pm0.10$ & 8.62& $\pm0.14$ & 8.67& $\pm0.14$ & $\times10^{-11}\,\mathrm{erg\,s^{-1}cm^{-2}}$ \\
\smallskip
$\chi_\mathrm{red}^2$/d.o.f. & 1.053&/386 & 1.100&/494 & 1.034&/589 & 1.042&/390 & 1.161&/333 & 1.167&/334 &  \\
\hline
\end{tabular}
\tablefoot{
\tablefoottext{a}{HR values as computed in Fig.\,\ref{fig:hr1}.}\\
\tablefoottext{b}{The additional model is the disk black-body. Temperature and radius of the black-body 
are relative to its inner boundary. The radius is expressed modulo $\sqrt{\cos i}$, where $i$ is the inclination of the system with respect to the 
line of sight.}\\
\tablefoottext{c}{Partial covering power-law model.}\\
\tablefoottext{d}{Normalization of the {\sc nthComp} or power-law component for HR$<$0.7 or HR$>$0.7.}\\
\tablefoottext{e}{Cross-calibration constants: $C1$ and $C2$ refer to obs1 and obs2. P and R indicate the \pn and \rgs.}\\
\tablefoottext{f}{The absorbed flux is computed from the \pn spectrum in the 0.5--11\,keV range of obs2. Uncertainties are only statistical.}\\
}
\end{center}
\label{tab:spec_hardness}
\end{table*}

\subsection{Pulse profile properties}
\label{sec:pulse_profile}
\begin{figure*}
  \begin{center}
	\includegraphics[angle=0, width=0.48\textwidth]{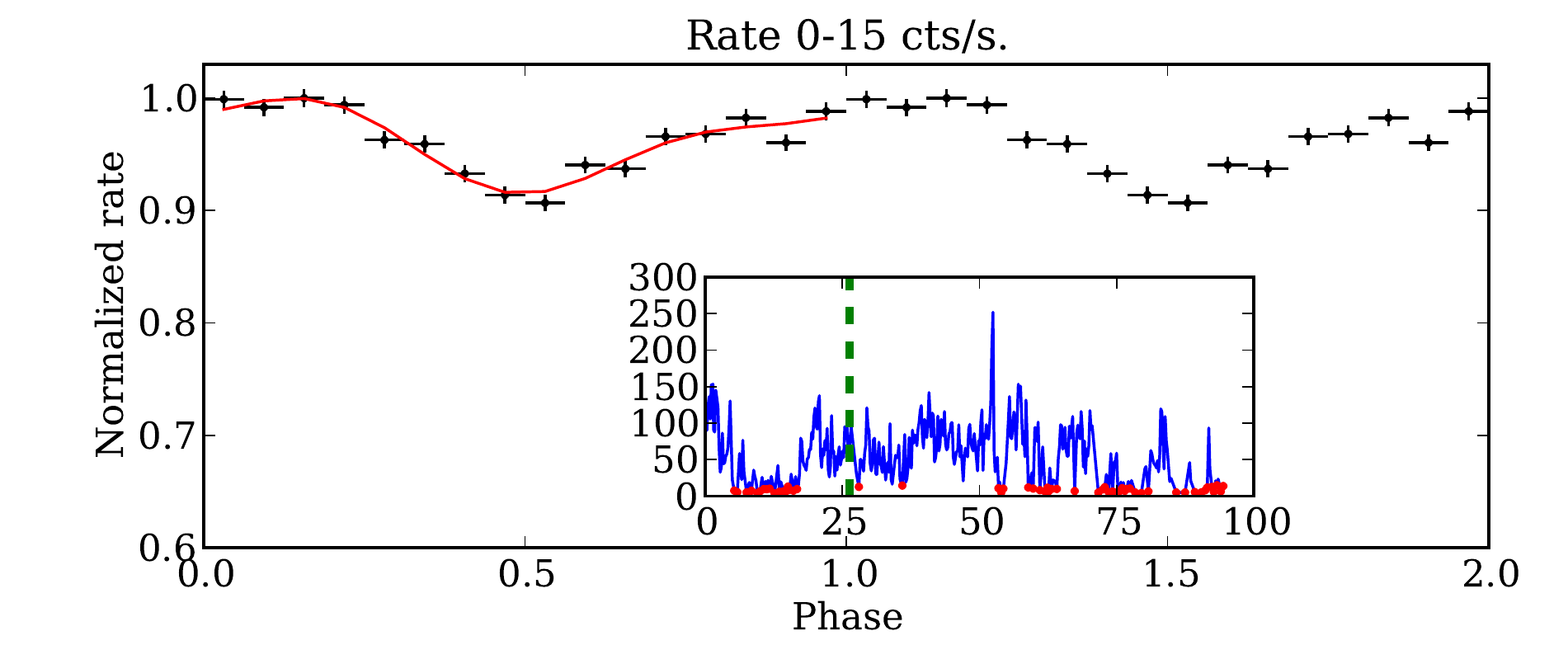}\hfill
      \includegraphics[angle=0, width=0.48\textwidth]{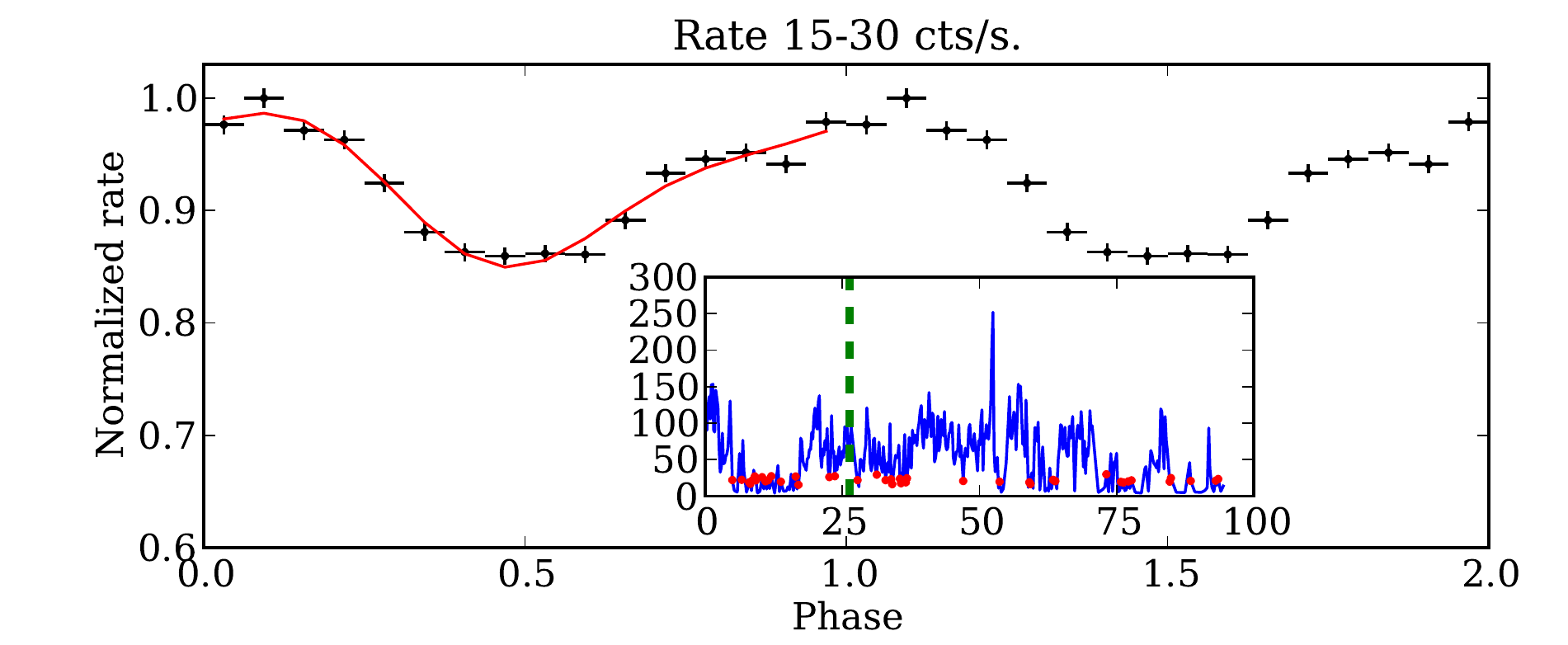}\\
	\includegraphics[angle=0, width=0.48\textwidth]{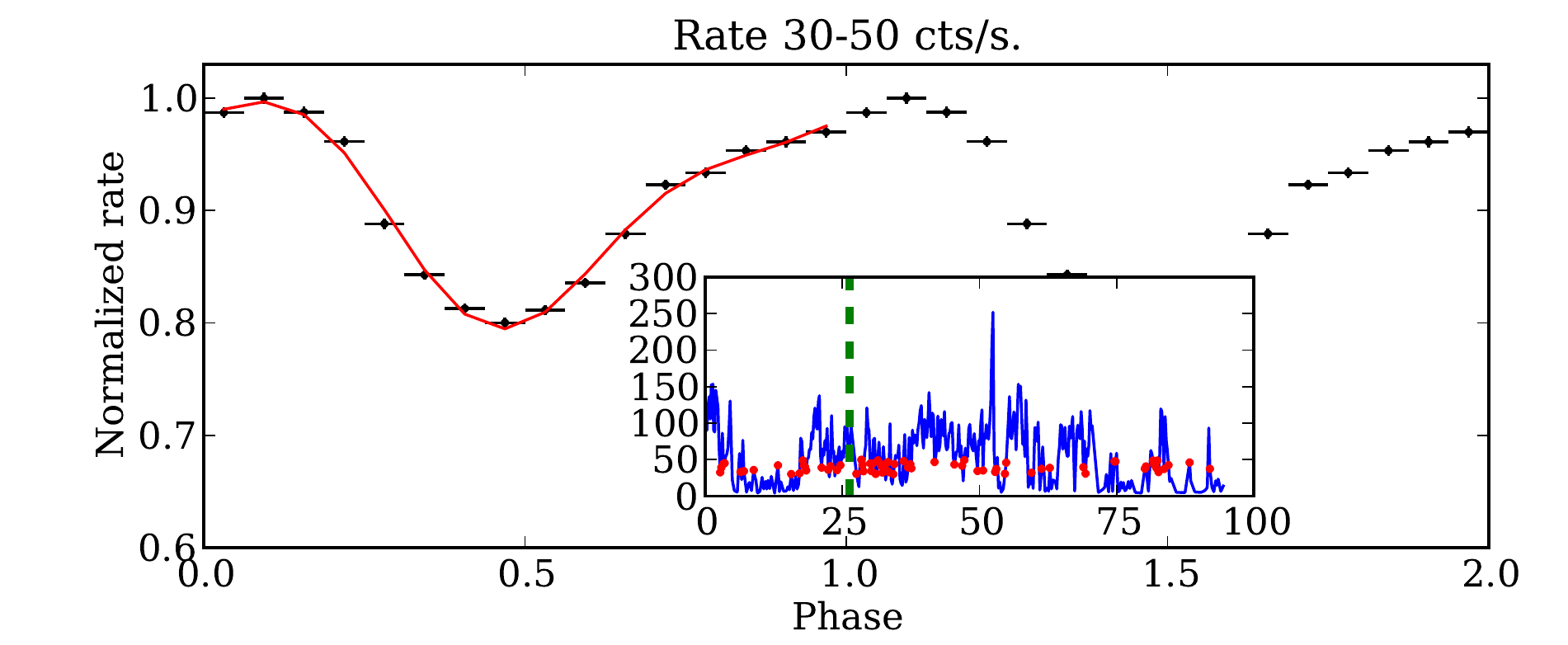}\hfill
      \includegraphics[angle=0, width=0.48\textwidth]{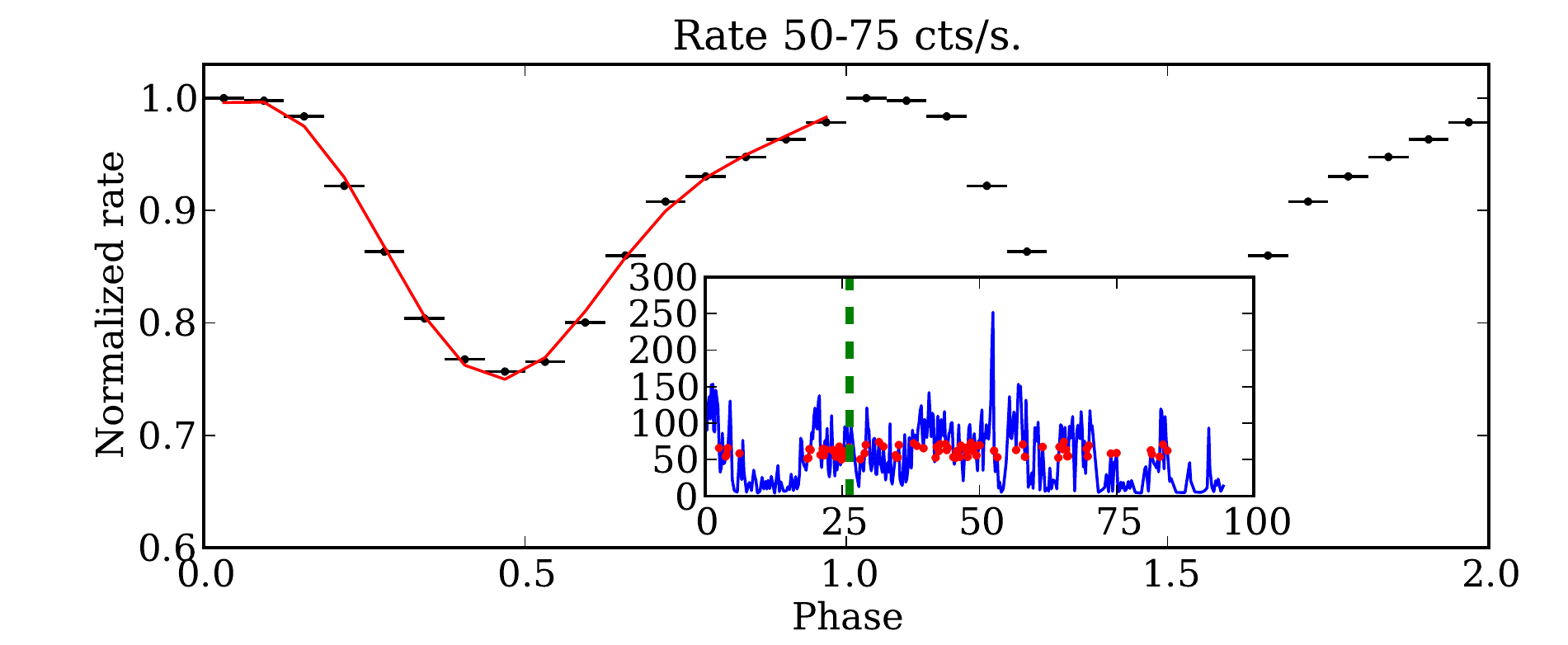}\\
      	\includegraphics[angle=0, width=0.48\textwidth]{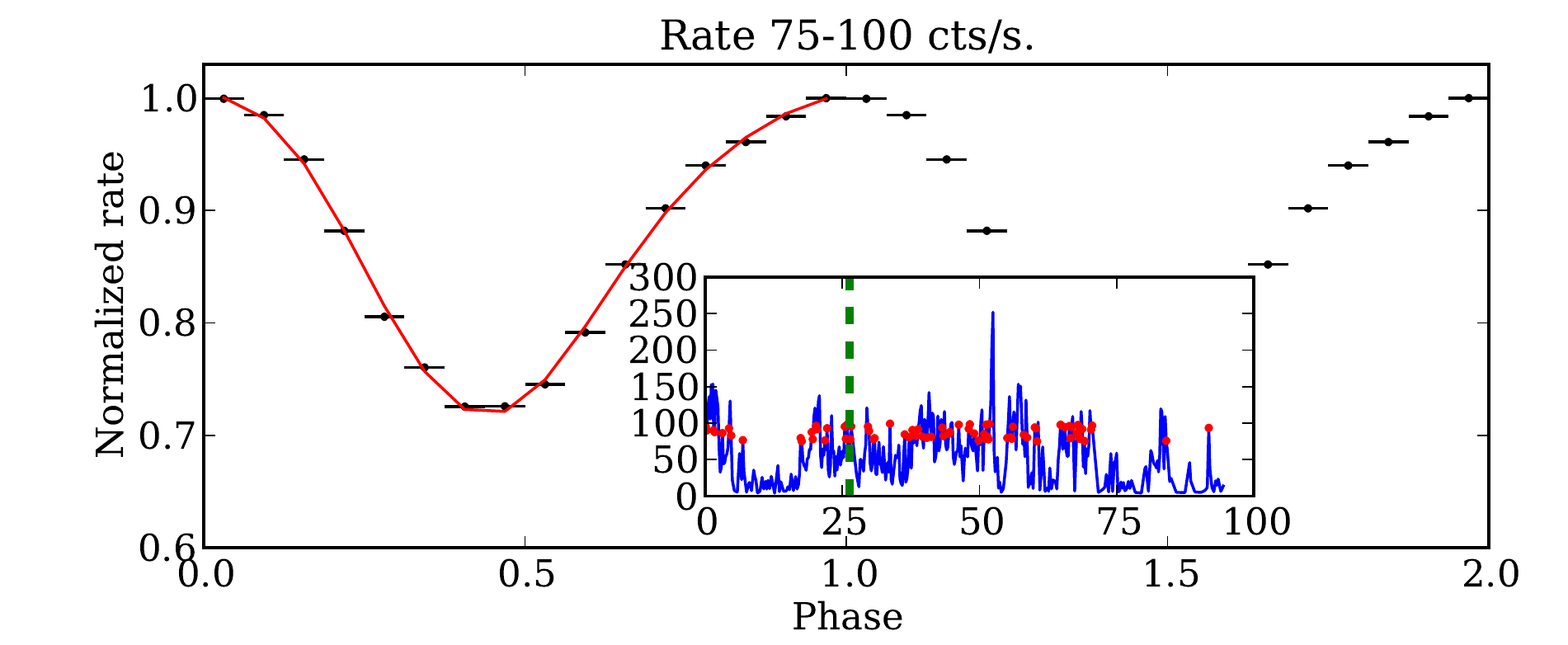}\hfill
      \includegraphics[angle=0, width=0.48\textwidth]{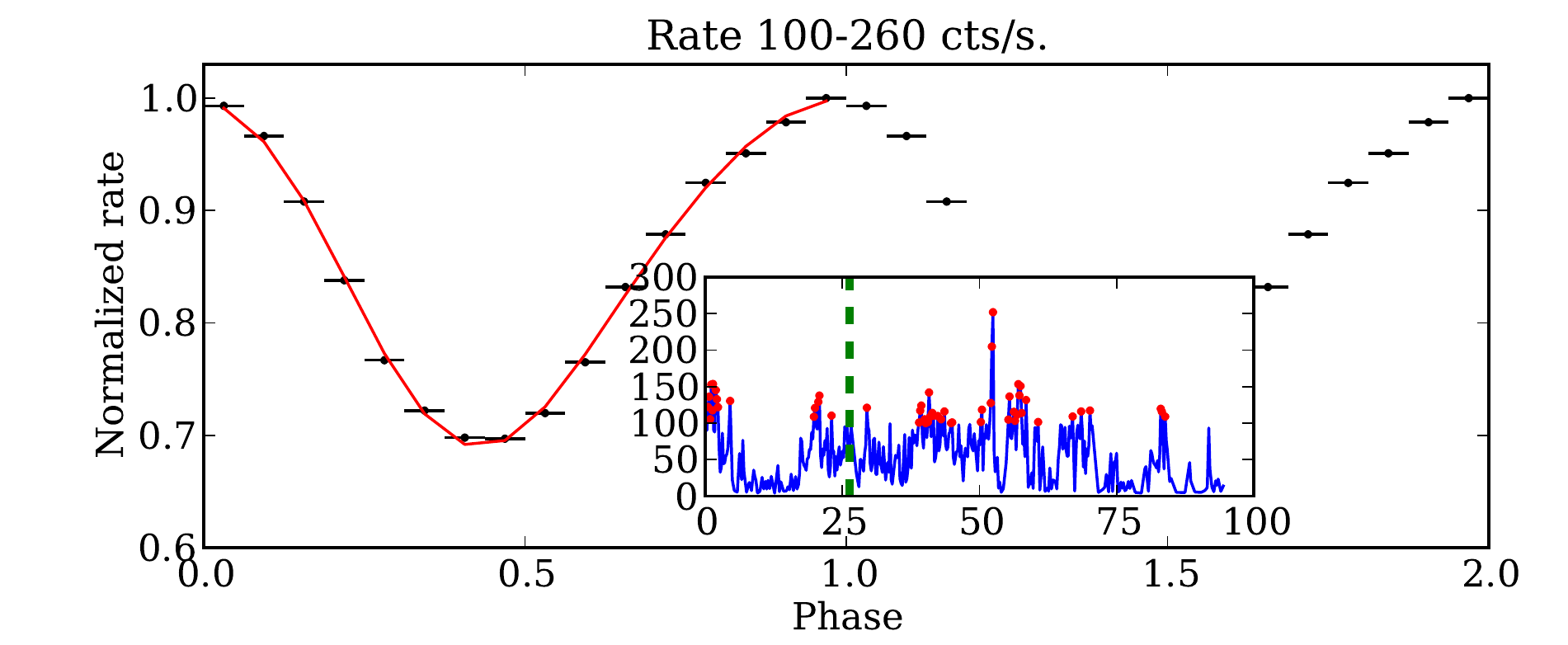}\\
	\caption{Pulse profiles of the combined obs1 and obs2 extracted during the different count-rate intervals of the blue states as in 
	Table\,\ref{tab:spec_rate}. In all panels of this figure, count rates were normalised 
	to the maximum. Pulses are repeated twice for clarity and the red solid lines represent the best fits to the profiles with two sinusoidal 
	functions as described in the text. The insets in each panel show the lightcurves of the two observations in the 0.5-11\,keV energy range (as in 
	Fig.\,\ref{fig:hr1}) with the time intervals used to extract the pulse profiles highlighted in red.} 
\label{fig:pulses}
\end{center}
\end{figure*}

To study the properties of the pulse
profile at different flux and energies, we first converted all the photon
arrival times into the barycenter of the system, using the
ephemeris given in paper\,I. Lightcurves were then folded at
the 3.9 ms spin period of the source, sampling each profile in 16
phase bins.  We investigated the source variability discussed 
in Sect.~\ref{sec:variability} in more detail 
by carrying out a count-rate- and HR-resolved
analysis of the pulse profiles.

\begin{figure}
  \begin{center}
	\includegraphics[angle=0, width=0.45\textwidth]{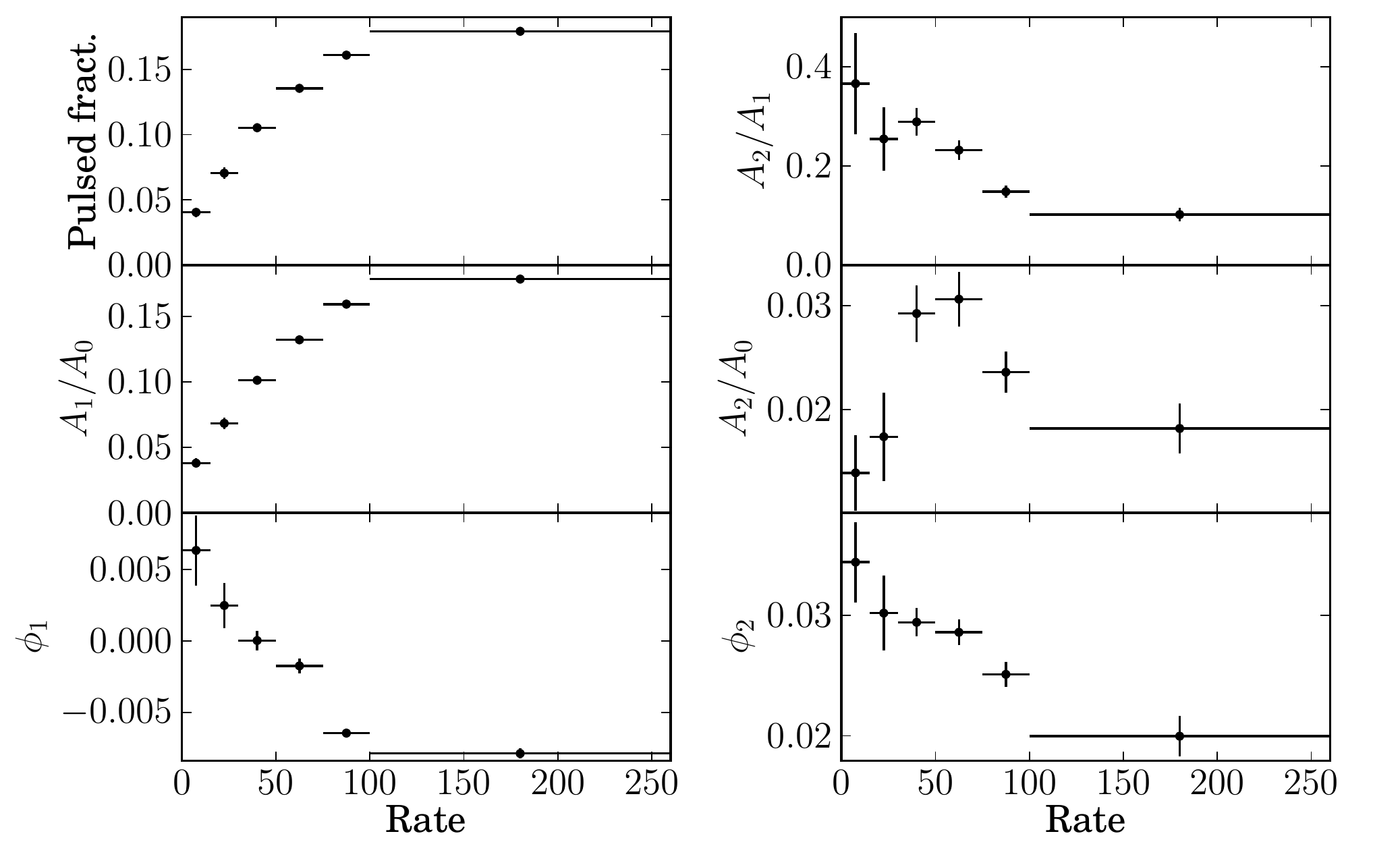}
	\caption{Parameters of the Fourier decomposition of the pulse profiles as a 
	function of the source count-rate in the \pn data.  
	The different parameters are defined in Eq.\,\ref{eq:fourier}. Uncertainties are obtained by 
	fitting eq. (\ref{eq:fourier}) to the pulse profiles and given at 1$\sigma$ c.l.} 
\label{fig:pulse_parameters}
\end{center}
\end{figure}

Fig.\,\ref{fig:pulses} shows the source pulse
profiles extracted during the same time intervals as were used for the
rate-resolved spectral analysis (see Table\,\ref{tab:spec_rate}). 
We fit these profiles by using a function comprising two Fourier
components:
\begin{equation}
p=A_0 + \sum_{n=1}^{2}A_n \cos (2\pi(n \phi - \phi_n))\,,
\label{eq:fourier}
\end{equation}
where $\phi$$\in$$(0,1)$ is the normalized phase. We verified that the
addition of higher Fourier components does not significantly improve
the fit. The best fits to the pulse profiles are shown with red lines
in Fig.\,\ref{fig:pulses}. The parameters determined from the fits are
shown in Fig.\,\ref{fig:pulse_parameters}. The pulsed fraction
$\sqrt{A_1^1+A_2^2}/A_0$ increases steadily from $4.0\pm0.2$\% to
$16.8\pm0.2$\% as a function of the source count rate.
The ratio of the amplitude of the
second and first harmonic decreases from $0.38\pm0.06$ to
$0.10\pm0.01$ as a function of the
count rate. Both the first and the second harmonics
underwent a remarkable phase shift at the highest count rates.

\begin{figure*}
  \begin{center}
	\includegraphics[angle=0, width=0.48\textwidth]{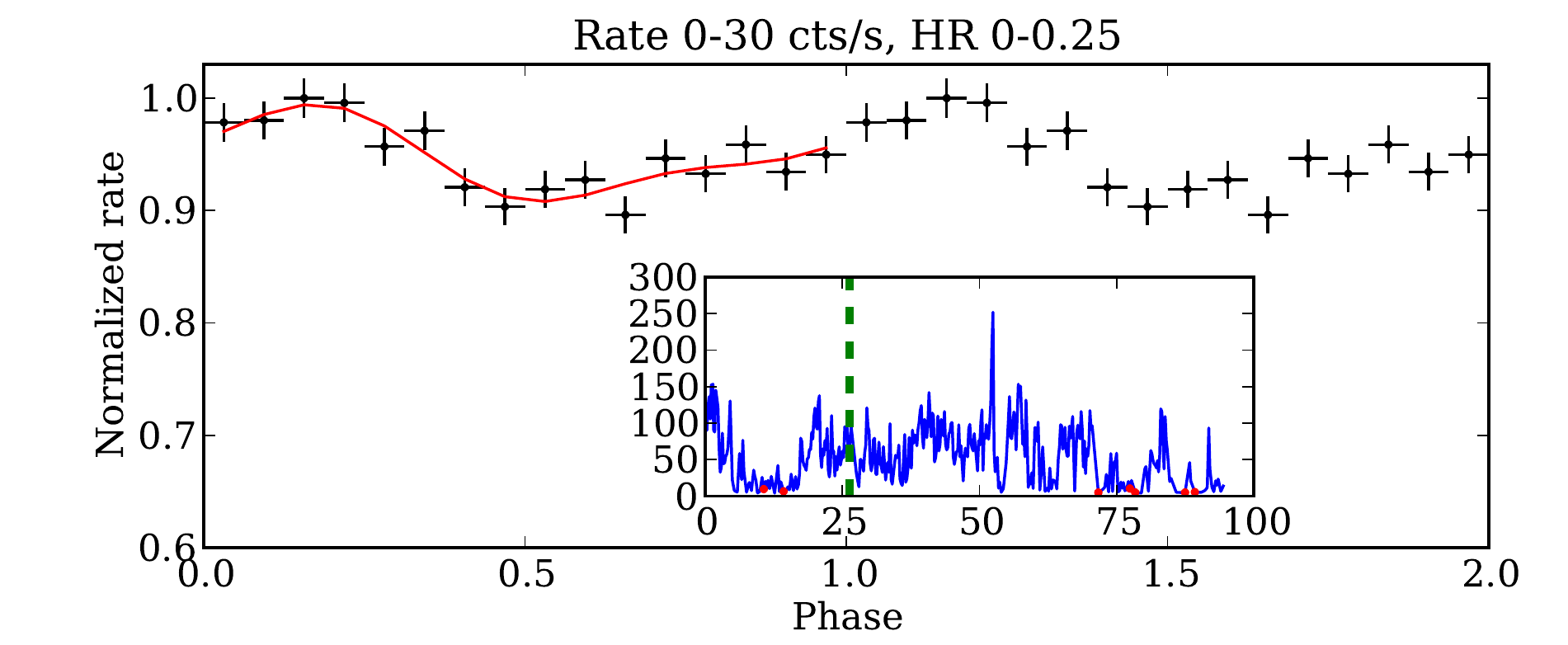}\hfill
      \includegraphics[angle=0, width=0.48\textwidth]{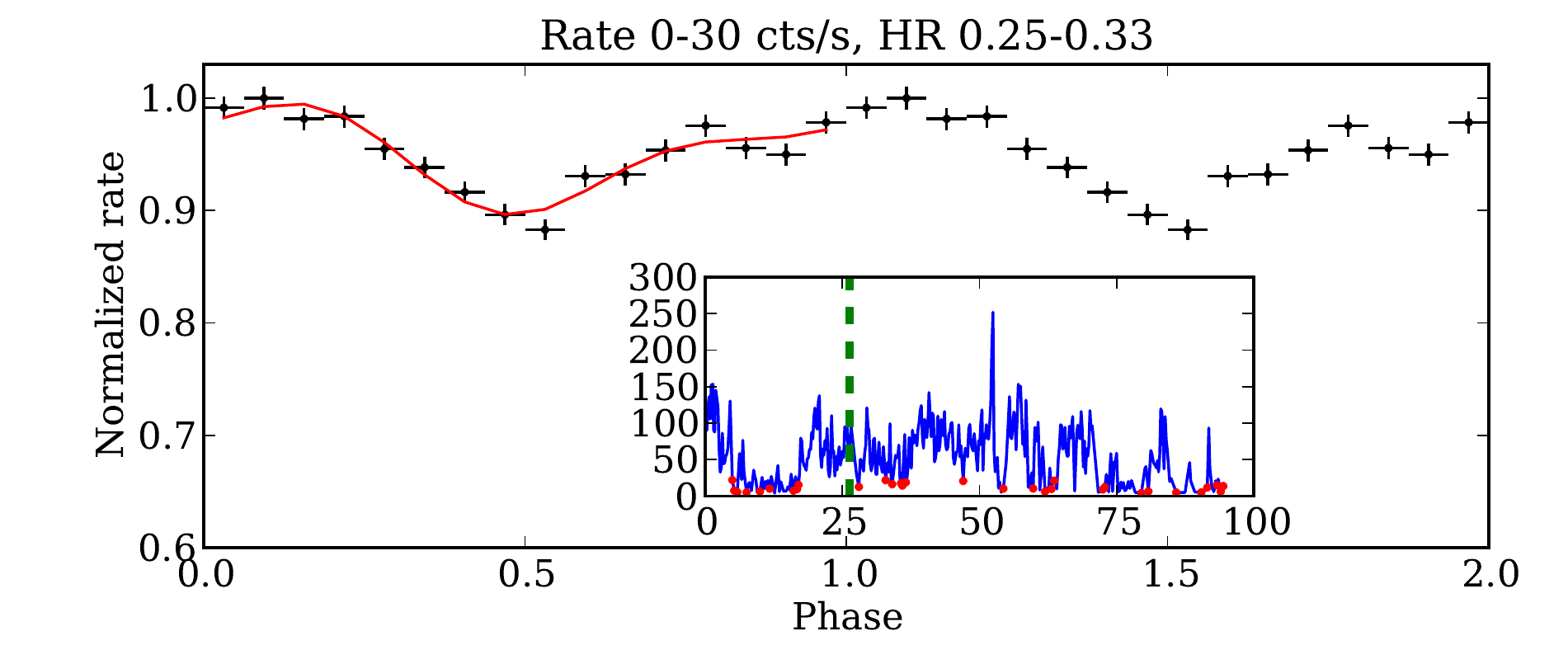}\\
	\includegraphics[angle=0, width=0.48\textwidth]{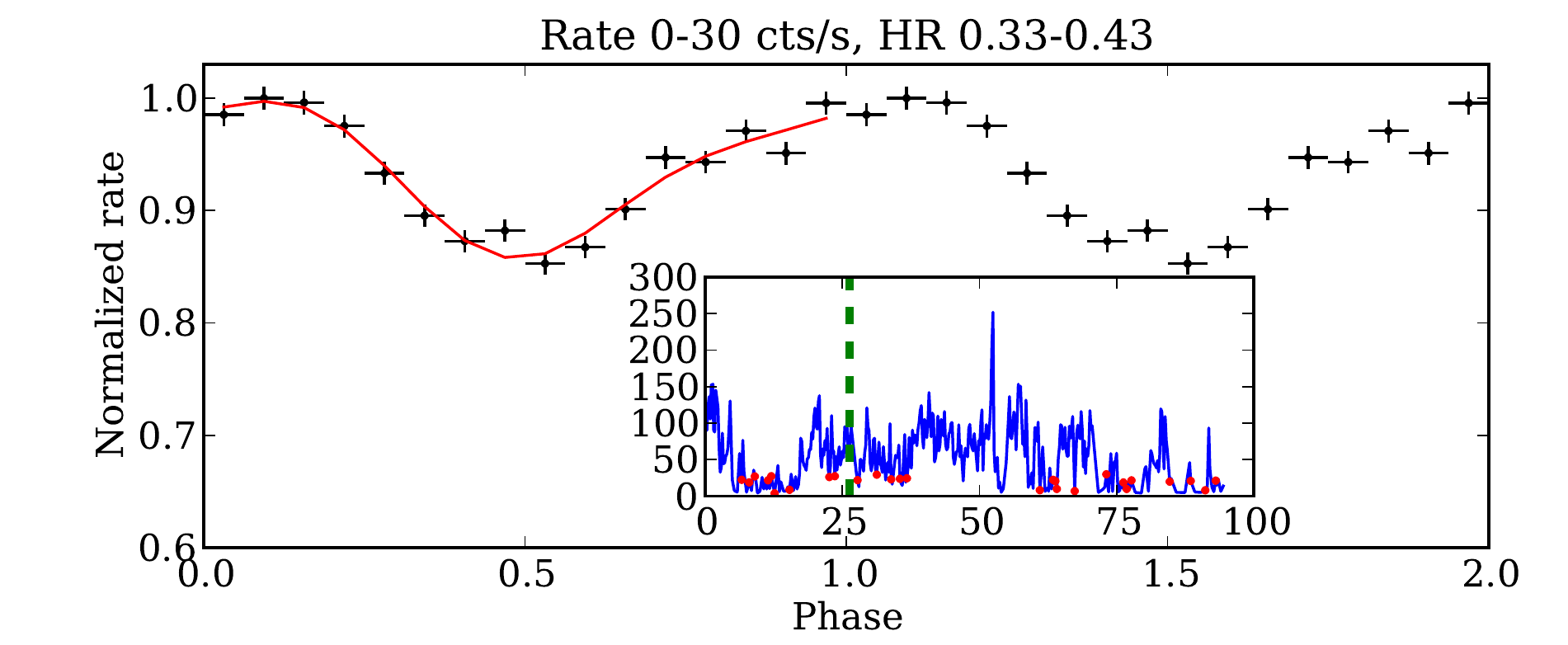}\hfill
      	\includegraphics[angle=0, width=0.48\textwidth]{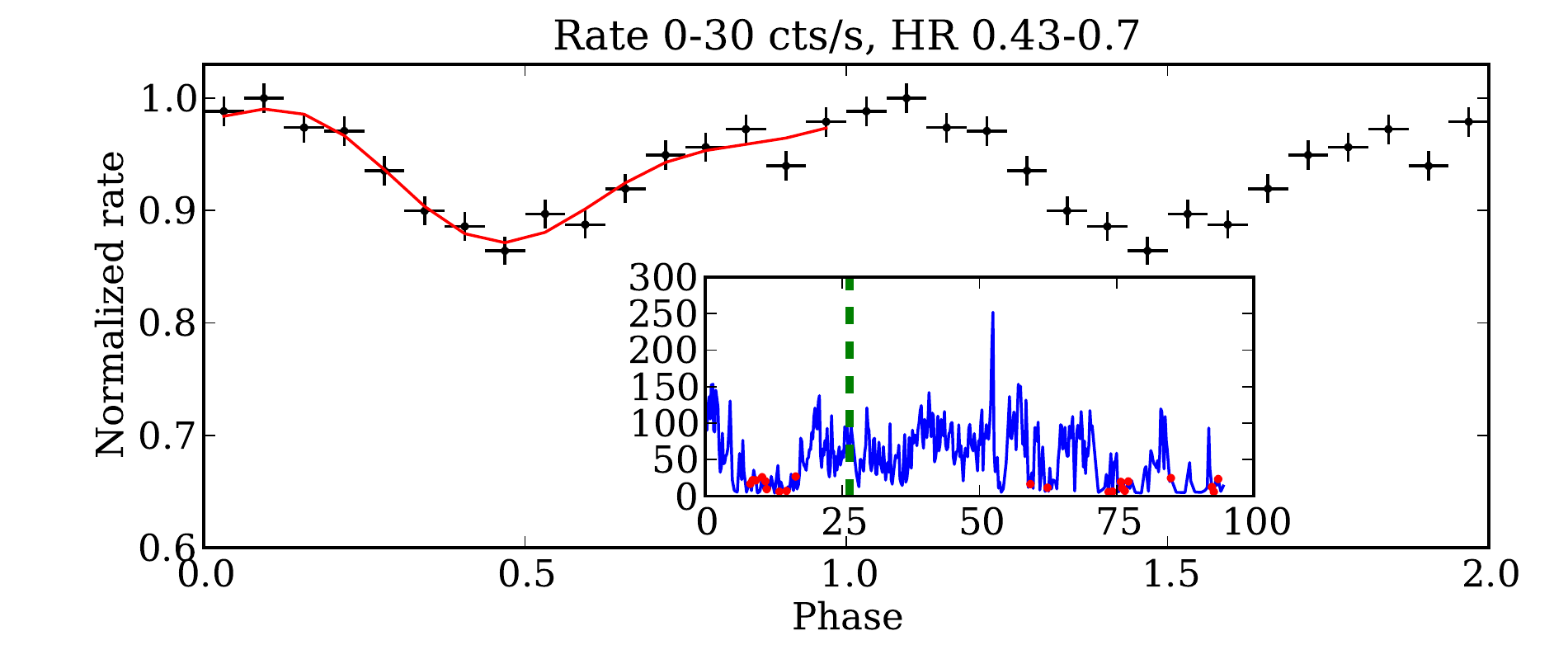}\\
      \includegraphics[angle=0, width=0.48\textwidth]{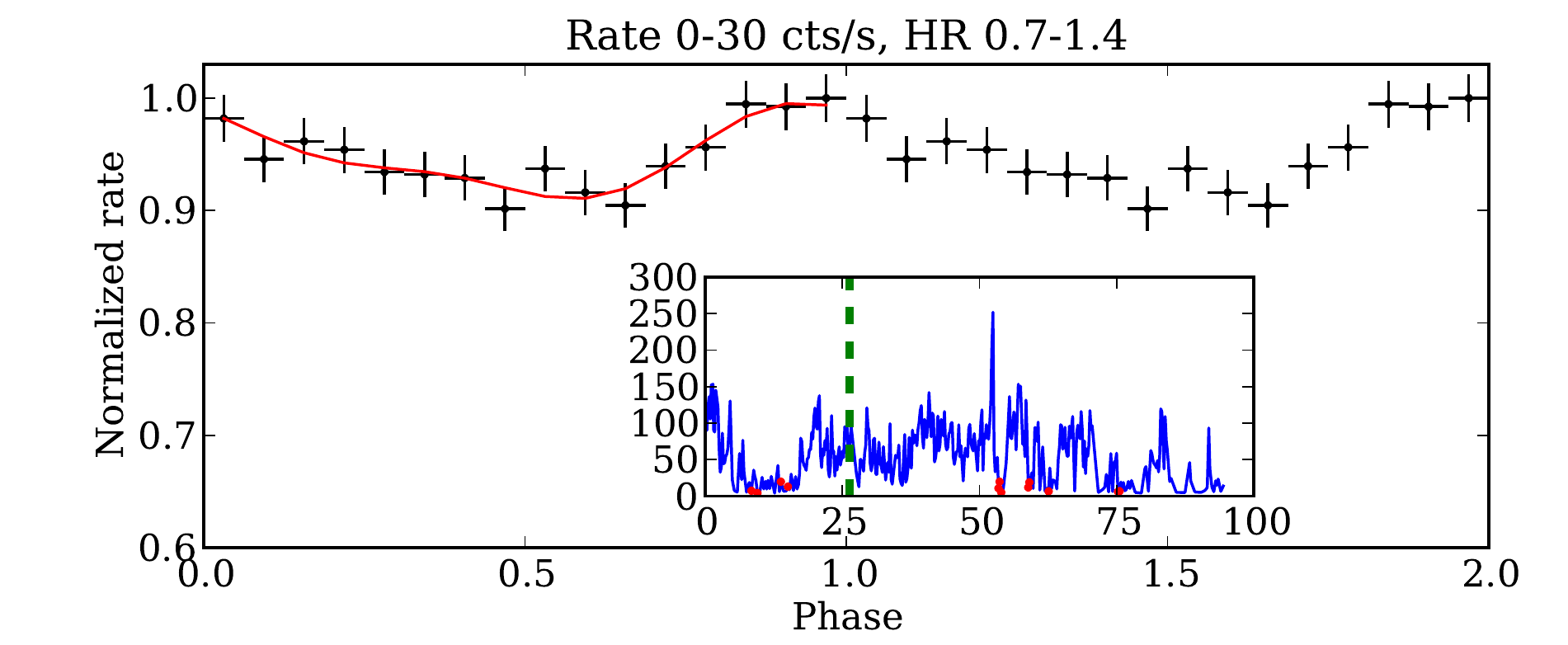}\\
	\caption{Same as Fig.\,\ref{fig:pulses}, but here the pulse profiles are extracted for different ranges
	 of the source HR in the magenta states.} 
\label{fig:pulses_hr}
\end{center}
\end{figure*}

\begin{figure}
  \begin{center}
	\includegraphics[angle=0,width=0.45\textwidth]{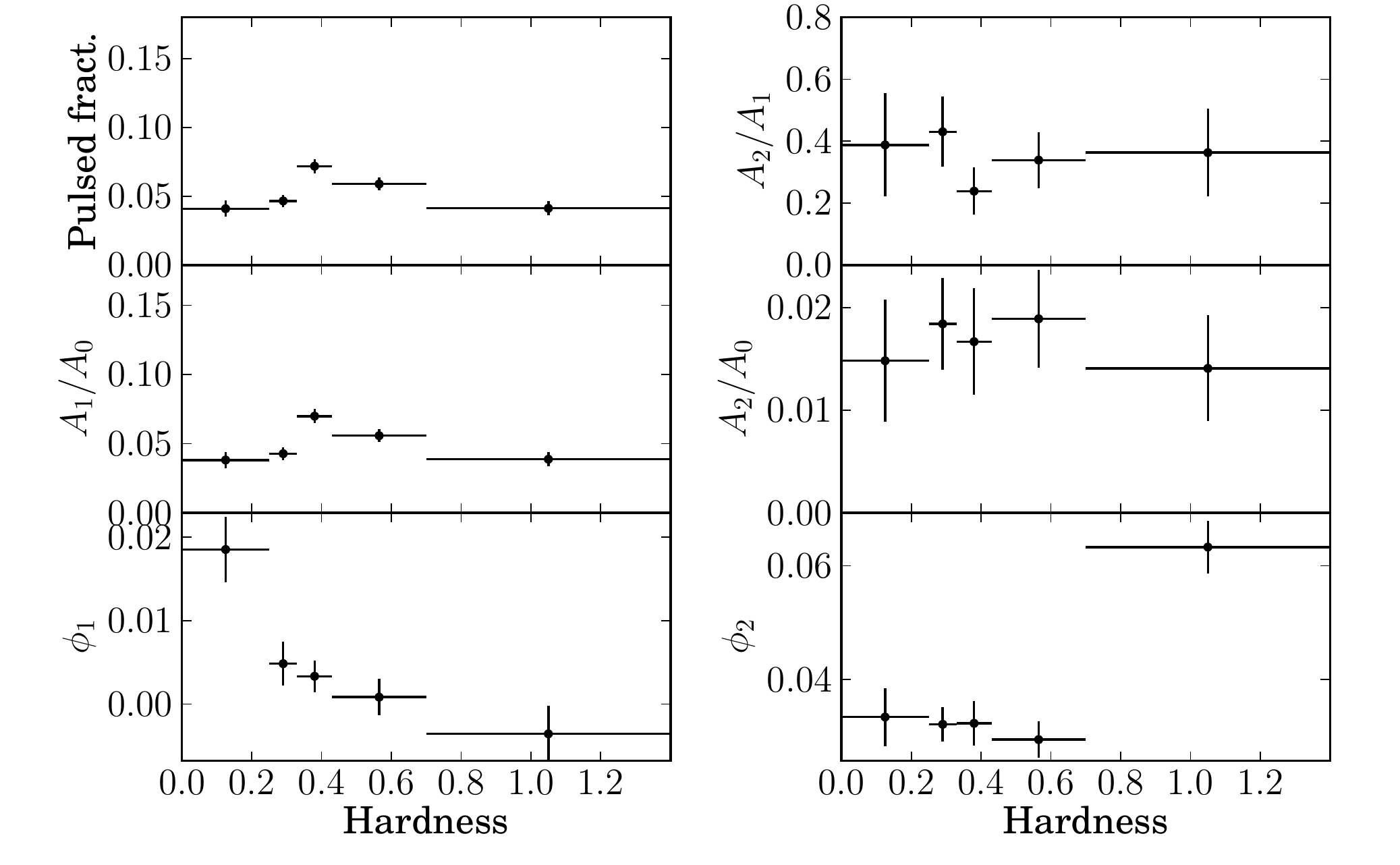}
	\caption{Same as Fig.\,\ref{fig:pulse_parameters}, but for the HR-resolved pulse profiles extracted in the magenta states 
	(Fig.\,\ref{fig:pulses_hr}). Uncertainties are obtained by fitting Eq. (\ref{eq:fourier}) to the pulse profiles 
	and given at 1$\sigma$ c.l.} 
\label{fig:pulse_parameters_hr}
\end{center}
\end{figure}

We carried out the same analysis as above for the pulse profiles extracted in the different HR intervals 
of the magenta states (see Table\,\ref{tab:spec_hardness}).  
Pulsations are significantly detected in all HR intervals, but the pulsed fraction was found to 
be as low as $\simeq5$\% (Figs.~\ref{fig:pulses_hr} and \ref{fig:pulse_parameters_hr}).
The relative contribution of the two harmonics remained constant as a function of the hardness. 
The phase of the fundamental decreased towards high HR and the first harmonic displayed 
a jump in the phase for HR$>$0.7, which shows a marked morphological evolution.

To summarize these findings and highlight the differences in timing properties between the magenta and blue states, 
we report in Fig.\,\ref{fig:hr_A1} the results 
of a timing analysis conducted for each time bin of Fig.\,\ref{fig:hr1} instead of by selecting 
specific intervals in the source count-rate and HR.   
In the panels (a) and (b) of this figure, we plot the pulsed fraction as a function of HR and total source intensity. 
Pulsations from the source are significantly detected in each time bin. The pulsed fraction
remains virtually constant at $\sim$5\% in the magenta states and increases steadily in the blue
branch.
We also computed 
the linear correlation coefficient for the different pulse profiles with respect to the average shape. 
This coefficient is plotted 
as a function of time and total source intensity in panels (c) and (d). 
From the comparison to the expected value of the linear correlation coefficient for identical 
pulses (the yellow band of panel (c)), it is clear that the pulse profiles 
drastically change their shape especially in the magenta states. 
We comment on these results in more detail in Sect.\,\ref{sec:discussion}.

\subsection{Aperiodic time variability}
\label{sec:aperiodic}
\begin{figure}
  \begin{center}
	\includegraphics[angle=0,width=0.45\textwidth]{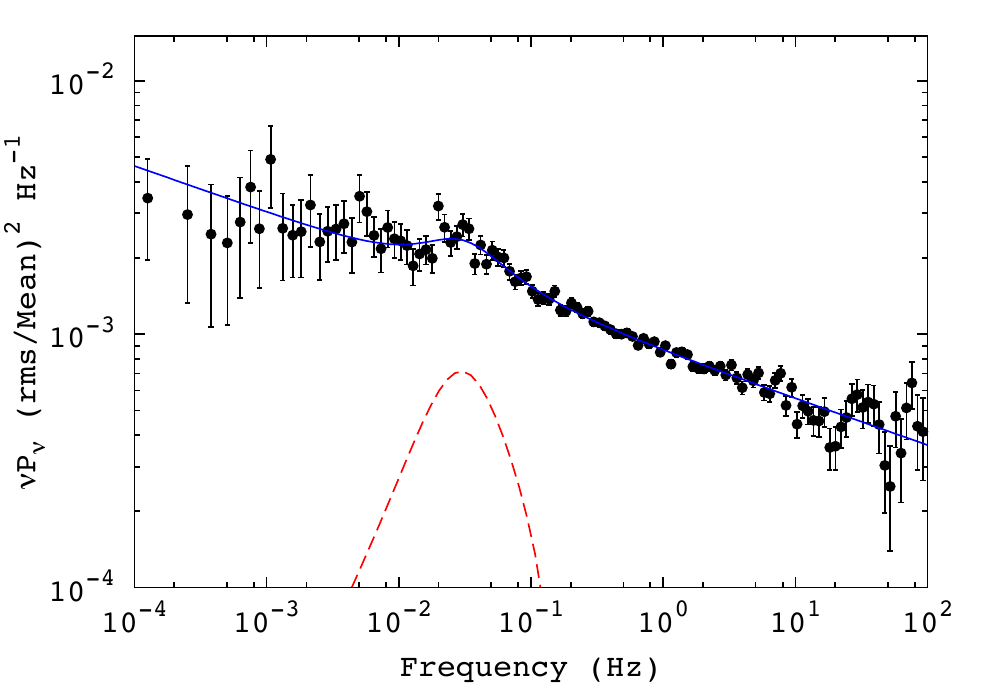}
	\caption{Power spectral density observed during obs2. The power spectrum was rebinned geometrically using a factor of 1.30, 
	and subtracted of a white noise level of 1.9948(5). The model used to fit the spectrum (red solid line) is the sum of a 
	flicker noise component (Eq.\,\ref{eq:pds}) and a Lorentzian (red dashed line).}
\label{fig:pds}
\end{center}
\end{figure}

We extracted the Fourier power spectra of \je in obs1 and obs2 by
  splitting them in eight intervals, lasting 3.3 and 7.9 ks,
  respectively, and averaging the power spectra evaluated in each of
  them. We checked that no variability took place on the time scale
set by the orbital motion and discarded from the analysis the
frequency bins corresponding to the first and second harmonic of the
pulsations.  The two power spectra were rebinned geometrically by a
factor of 1.1.  The aperiodic variability of \je is
characterized by a power-law noise $P(\nu)\sim \nu^{-\beta}$ with
$\beta\simeq1.2$ (i.e., flicker noise).  The addition of a
flat-top noise component significantly improves the fit of both
power spectra, which we modelled with the function:
\begin{equation}
P(\nu)=WN+C{\nu}^{-\beta}+\frac{N}{1.+(2 \nu/W)^2}.
\label{eq:pds}
\end{equation}
This aperiodic noise component has an RMS fractional amplitude in
the band between $10^{-4}$ and $100$\,Hz that
exceeds 90\% in both observations.  No other features were
detected, even when averaging over shorter time intervals and/or using
different rebinning factors. We also searched for changes of the power
spectral shape at different energies, but found no significant
variations with respect to the spectrum extracted over the entire
0.5--11 keV energy band. As an example, we show the power spectral
density of obs2 in Fig.\,\ref{fig:pds} together with the best-fit
model.  The best-fit parameters to the power spectra densities of obs1
and obs2 are given in Table\,\ref{tab:psd}.

In obs1, where the separation between the magenta and blue states is clearer, we carried out 
a power spectral density (PSD) analysis separately for the two states.
We found that the functional shape of the PSD remained unchanged between the two states, but the characteristic 
frequency of the variability decreased in the magenta states.  
This can be also inferred also by visually inspecting  
the lightcurve. PSD fitting reveals that $\beta$ 
increases from 1.11$\pm$0.02 to 1.27$\pm$0.02 when passing from the blue to the magenta states, while
the flat-top component remains unvaried within uncertainties.

\begin{table}[h!]
\caption{Best-fit model parameters of the power spectra density extracted by using the entire exposure time available 
in obs1 and obs2.} 
\begin{center}
\begin{tabular}{lll}
\hline
\hline
   & obs1 & obs2 \\
\hline
WN & $1.994(2)$ & $1.992(1)$\\
C & $1.92(4)$ &  $2.01(2)$ \\
$\beta$ &  1.20(1) & 1.18(1) \\
N & $107\pm32$ & $73\pm20$ \\
W & $0.08(1)$ & $0.048(6)$ \\
$\chi^2_r$ (d.o.f) & 1.33 (105) & 1.07 (114) \\
\hline
\end{tabular} 
\end{center}
\label{tab:psd}
\end{table}

\section{Discussion}
\label{sec:discussion}

We reported on an in-depth analysis of the two
\xmm\ observations carried out in the direction of \je\ during its first observed 
X-ray outburst. 

We first studied the source average X-ray spectrum and showed that the overall spectral properties of the 
X-ray emission from \je\ are qualitatively similar to those of other AMXPs in outburst 
\citep[Fig.\,\ref{fig:spec_all},][]{poutanen2006, gilfanov1998,gierlinski2005}. 
Compared with the results we reported in paper\,I, including the \rgs spectra in this work allowed us to better 
constrain the apparent position of the inner accretion disk radius. We found that $r_\mathrm{in}$$\simeq$66-77$\sqrt{\cos i}$\,km 
for an assumed distance to the source of 5.5\,kpc. The actual inner radius can be up to a factor $\sim$2 larger when color corrections 
are considered \citep{merloni2000} and a torque-free inner boundary condition is applied \citep{gierlinski1999}.

At odds with all the other AMXPs in outburst observed so far \citep[see, e.g.,][for a
recent review]{patruno12}, \je\ displayed a very peculiar variability in the X-ray domain and underwent  
remarkable switches between low- and high-intensity states during both the \xmm\ observations.  
The two preferred states were first noticed in the histogram of the 
source count-rate, which that satisfactorily described only by using two log-normal distributions. 
The first of the two distributions, marked in magenta, dominates the contribution to the total source count-rate lower than 30\,cts/s 
in the \pn 0.5--11\,keV band. The second distribution, marked in blue, provided the largest contribution at higher count rates. 
By performing a more detailed analysis on the source light curves, we showed that  
the HR showed limited changes ($\sim$20-30\%) around a value of $\sim$0.5 in the blue state, 
which suggests little spectral variability. 
During the magenta state, the source intensity was found to be on average 
a factor of ten lower than in the blue state and characterized 
by prominent swings in the HR (see Sect.\,\ref{sec:results}). 
This indicates that significant spectral changes take place 
in the magenta states. 
To investigate the origin of the rapid variability, we therefore carried out a count-rate-resolved spectral and timing analysis of the blue 
states, and an HR-resolved spectral and timing analysis of the magenta states. The results of these analyses are discussed below.

In the blue state, the spectral continuum is well described by a thermal
Comptonization component plus black-body emission from an accretion disk. 
Both components remain virtually constant across obs1 and obs2, 
displaying consistent parameters at different source intensities (we only measured a slight hardening in the photon-index 
of the Comptonization component, decreasing from $\Gamma$$\simeq$1.6 to $\Gamma$$\simeq$1.4 at the highest count rates; see 
Table\,\ref{tab:spec_rate}). In particular, no significant changes are recorded 
in the equivalent width of the iron line and the absorption column density. 
These results disfavor any interpretation of the prominent variability of the source intensity observed 
during the blue state in terms of absorption dips and suggest, instead, that it is due to intrinsic variations of 
the mass accretion rate onto the NS.
These variations regulate the position of the disk inner boundary (see Eq.~\ref{eq:rm}), pushing it towards the 
NS surface at high luminosity and enhancing the separation between the inner rim and the NS surface at lower luminosity.
By invoking this mechanism, the observed strong correlation of the source 
pulsed fraction with its X-ray intensity can be explained as it follows: the antipodal spot might become visible,
which leads to a reduction of the pulse fraction and an increase of the harmonic content
\citep{beloborodov2002,ibragimov2009,kajava2011,riggio2011},  an annular-shaped accretion stream footprint might shrink
\citep{ibragimov2011}, or variations of the spot latitude might be present \citep{lamb2009}.
Unfortunately, the constraints on $r_{\rm in}$ derived from the different 
rate-resolved spectra in the blue states are too weak to provide a firm measurement of this effect  
(see Table~\ref{tab:spec_rate}).    

Fits to all the spectra extracted in the magenta state revealed a slight decrease of the  
absorption column density, which becomes  compatible with the Galactic value 
in the direction of the source \citep[$2.4\times10^{21}\,\mathrm{cm}^{-2}$;][]{becker2003}.
This is possibly related to the presence of less accreting material 
in the surrounding of the neutron star. The thermal component due to the accretion disk 
observed in the blue state was no longer detectable. Instead, a significantly hotter ($\sim$0.4-0.7\,keV) 
and more confined (few km) thermally emitting region was revealed during the spectral fits. 
Similar regions are usually associated to hot spots on the NS surface \citep[][and references therein]{gierlinski2002,gierlinski2005,papitto10}. 
To verify that the accretion disk component was not gone undetected only because of the reduced statistics of the spectra 
extracted during the magenta states, we added to the relevant fits a disk black-body component with a temperature fixed 
to the value measured in the blue state.  We derived upper limits on $r_\mathrm{in}\sqrt{\cos i}$ between 
9-42\,km (at 90\% c.l), which demonstrates that, if present, such a disk should have been detected in the magenta spectra. 
These results agree with the scenario proposed above for the blue states. At the lower fluxes that characterize 
the magenta states, the inner boundary of the disk recedes farther away from the neutron star surface 
and thus does not produce a significant contribution to the source X-ray spectrum. The hot surface of 
the neutron star becomes visible along the observer's line of sight and dominates 
the soft X-ray emission from the source. 

\begin{figure*}
  \begin{center}
	\includegraphics[angle=0,width=0.9\textwidth]{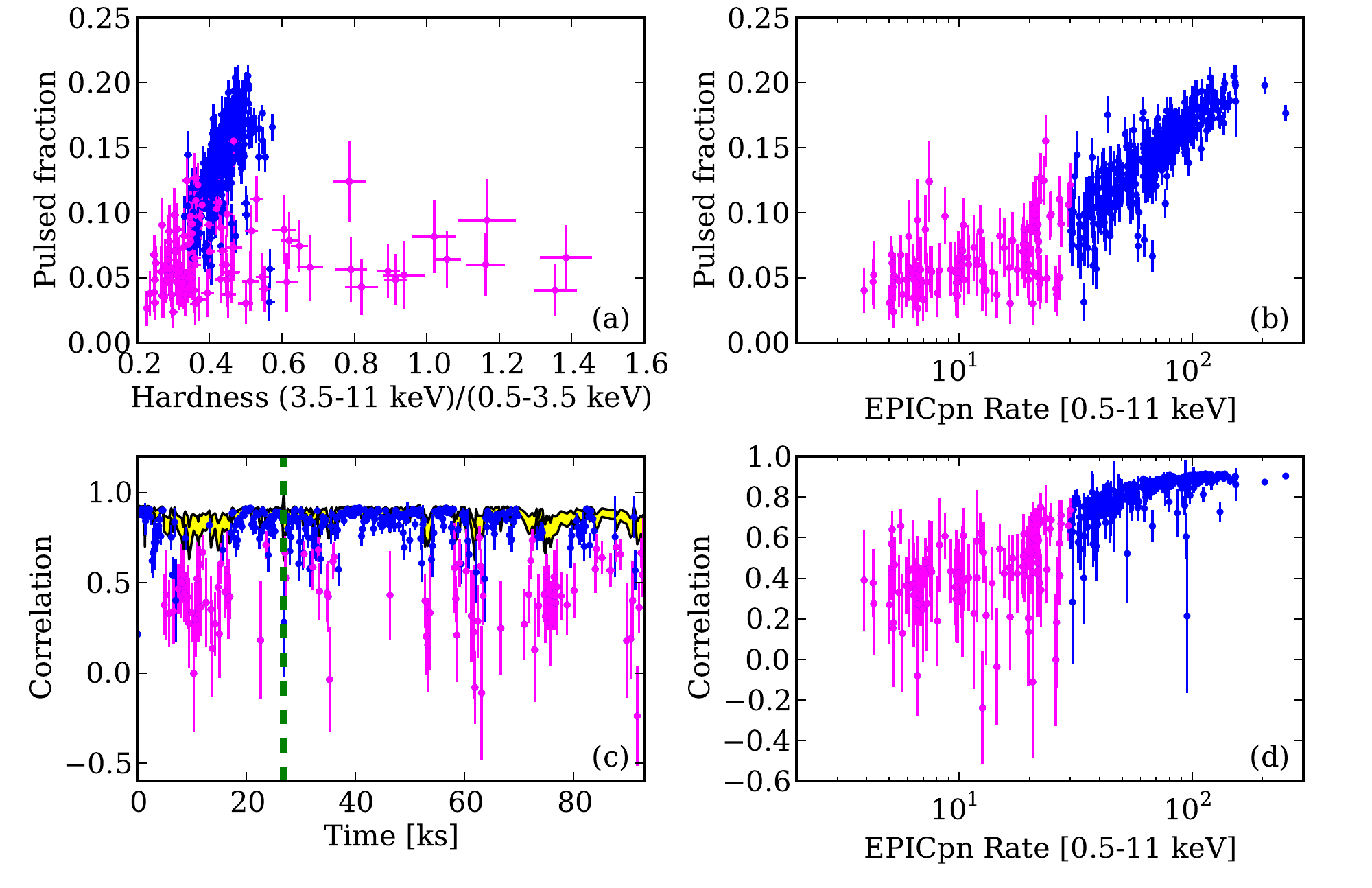} 
	\caption{Timing analysis performed for every time bin with length of at least of 200\,s and S/N$\ge$25 in obs1 and obs2. The origin of times is the same 
	as specified in Fig.\,\ref{fig:hr1}. {\it Panel (a)}: pulsed fraction as a function of the HR. 
	  {\it Panel (b)}: pulsed fraction as a function of the source total intensity. {\it Panel (c)}: linear correlation coefficient between the average pulse 
	  profile and the pulse profile computed in each time bin 
	  (the shaded yellow area represents the expected range of correlation values at 1$\sigma$ c.l. that all profiles should have 
	  if no changes occur other than the average one; the level of timing noise of the observation was also taken into account).  
	  {\it Panel (d)}: the correlation coefficient as function of the source count rate.
	  The uncertainties are reported  at 1$\sigma$ c.l. and are computed for the pulsed fraction and correlation coefficient
	  using a bootstrapping technique with 1000 realizations.}
\label{fig:hr_A1}
\end{center}
\end{figure*}

When the inner boundary of the accretion disk recedes beyond a critical value, the so-called propeller effect  
is expected to set in \citep{illarionov75}. In this regime, accretion onto the neutron star is (at least) partially 
inhibited by the rapid rotation of its magnetic field lines, and 
major changes in the source's pulsed fraction, pulse profiles, and spectral shape are expected. 
Interestingly, the pulse fraction of \je\ reaches its minimum in the magenta states 
(see Figs.~\ref{fig:pulses_hr} and \ref{fig:pulse_parameters_hr}) and the corresponding 
pulse profiles changes drammatically. The latter are nearly sinusoidal in the blue states and become 
more and more distorted at the lower source intensities (see Figs.~\ref{fig:pulses}, \ref{fig:pulse_parameters}, and \ref{fig:hr_A1}). 
Phase shifts are also clearly detected and observed to increase as a function of the HR (Fig.~\ref{fig:pulse_parameters_hr}).  
Similar phase shifts can occur in accreting systems when (i) the magnetic field footprints on the 
compact star surface migrate as a consequence of changes in the pattern followed by the inflowing material; (ii) 
there exists a difference between the phase of the black-body and Comptonized emission; (iii) the beaming of the radiation 
from the accretion column changes as a function of the source intensity 
\citep[see][for a discussion and further references]{ibragimov2009}.
These findings thus suggest that accretion in the magenta states is significantly reduced and takes place through complex patterns, 
as expected if a propeller-like mechanism regulates 
the X-ray emission and variability of the source.

When inspected as a function of the HR, both the spectra and the pulse profiles of the magenta states displayed 
a particularly intriguing change for HR$>0.70$. A striking comparison is shown in Figs.~\ref{fig:spec_lowHR} and \ref{fig:spec_highHR}.
Correspondingly, a strong jump in
phase of the overtone is also observed (Fig.~\ref{fig:hr_A1}).   
The spectrum extracted during the time intervals when HR$>0.70$ could not be well described 
by a thermal plus a Comptonization component. We showed that a  
partially covering power-law model provided a significantly better fit to the data. 
The additional absorbing material implied by this model 
requires that part of the power-law emission from the source is seen through 
an absorption column density that is a factor of $\sim$10 higher than the average value. 
We suggest below that the increased column density of the hard magenta states 
can be associated with the production of strong outflows during the switch from a 
weak to the strong propeller regime.

The definition of weak and strong propeller was introduced by \cite{romanova04} following the results of  
2D and 3D MHD simulations of accreting AMXPs. If we estimate the position of the inner boundary
of the accretion disk around the NS with the usual equation
\citep[see, e.g.][]{frank03}
\begin{equation}
R_{\rm m} = 3\times10^6 L_{36}^{-2/7} \mu_{26}^{4/7} {\rm cm}
\label{eq:rm}  
\end{equation}
(here $L_{36}$ is the X-ray luminosity in units of 10$^{36}$\,erg/s and $\mu_{26}$ is the NS magnetic moment in units of 10$^{26}$\,G\,cm$^3$),  
it is well-known that accretion onto the NS can proceed unimpeded as long as this radius is smaller than the so-called corotation radius
\begin{equation}
R_{\rm co} = 1.5\times10^6 P_{-3}^{2/3} {\rm cm}.
\label{eq:rco} 
\end{equation}
Here $P_{-3}$ is the NS spin period in units of 1\,ms and we assumed
for both $R_{\rm co}$ and $R_{\rm m}$ an NS mass of 1\,$M_{\odot}$ and
a radius of 10\,km.  The corotation radius represents the distance
from the NS at which the rotational velocity of the NS magnetic field
lines, anchored to the star, is higher than that of the material at the
inner boundary of the disk. When $R_{\rm m}$ approaches
$R_{\rm co}$, accretion is partly inhibited by the fast rotation of the
star and material at the inner boundary of the disk tends to be
ejected from the system instead of being accreted.  
In the weak propeller, the magnetospheric radius is relatively close to the
corotation radius ($R_{\rm m}$$\simeq$$R_{\rm co}$), and for typical parameters the 
quantity of material that is accreted onto
the star is comparable with the amount that is ejected through
outflows \citep[see Figs.\,14-16 in][]{ustyugova06}. 
The latter can be in the form of winds emerging from the
external layers of the accretion disk or collimated in mildly relativistic
jets. When $R_{\rm m}$$\gtrsim$$R_{\rm co}$, the
system switches to the strong propeller regime. In this regime, the
amount of ejected material in the outflows can be as high as 10 times
the one that is effectively accreted onto the NS and up to $\sim$70\%
of the total mass inflow rate along the disk \citep[see also][and references therein]{lii12,lii13,lovelace13}.  
The material ejected from the system through the outflows attains relativistic speeds and
thus a number of different physical processes (e.g., Fermi
acceleration caused by shocks between the ejected material and the
surrounding medium, and inverse Compton scattering of low-energy
photons by the bulk motion of the outflow) are expected to contribute
to the source X-ray emission and lead to a harder
spectrum than that produced by the accretion. A similar
scenario was also invoked to explain the hard spectra of the
intermittent AMXP Aql\,X-1 during its transitions from accretion to
the propeller state \citep{zhang98} and in similar transitions
observed in a number of accreting pulsars \citep[see e.g.,][and
references therein]{cui97}. The presence of outflows in \je\ due to possible 
strong propeller phases is also supported by the detection of a very pronounced 
variability of the source in the radio domain 
\citep[see Appendix\,\ref{sec:appendix2} and][]{pavan13}.

Assuming an average X-ray luminosity of
10$^{35}$~erg/s for all the magenta states, the condition $R_{\rm m}$$\simeq$$R_{\rm co}$ for the
onset of the weak propeller in \je\ would imply a magnetic field for
the NS in this system of $\sim$3$\times$10$^8$\,G, which is in the
range usually inferred for other AMXPs \citep[see, e.g.,][and
  references therein]{papitto12} and consistent with the value given
previously in paper\,I.  However, it is well-known that large
uncertainties affect the estimates of the magnetospheric radius given
in Eq.\,(\ref{eq:rm}), and therefore the NS magnetic field inferred from
this argument should be taken with caution \citep[see discussion
 in][]{bozzo09b}.  Limits on the NS magnetic field can also be inferred
 from \je's radio pulsar nature. For the NS to be active
as a radio pulsar, the electrical voltage gap (formed by a dipolar magnetic field that 
corotates with the NS) should
extract charged particles from the NS surface, accelerate them along
the magnetic field lines, and eventually lead to a pair cascade that is
ultimately responsible for the observed coherent radio emission
\citep{Goldreich1969,Ruderman1975}.  The so-called pulsar death
line sets a theoretical limit for this process to take place. In a
purely central dipolar case, for the spin period of \je, this would lead
to a lower limit of the magnetic field of B$>6.5\times10^{7}$\,G
\citep[see Eq.~(6) in][]{Chen1993}\footnote{Note, however, that
  when different configurations of the magnetic field
  (i.e. twisted field lines) are considered, this field limit can decrease by about
  an order of magnitude \citep[$\sim5\times10^{6}$\,G; see Eq.~(9)
    in][]{Chen1993}.}.  Furthermore, we note that the weakest magnetic
field ever measured for a radio pulsar is 6.7$\times10^{7}$\,G for the
binary millisecond pulsar PSR J2229$+$2643 \citep[spin period
  2.9\,ms;][]{Camilo1995,Wolszczan2000}.  An upper limit can be
determined from the pulsar's spin-down rate during the
rotationally-powered phase.  However, we note that a strong limitation
to this measurement is provided by the motion in the cluster's
gravitational field, which yields a spurious frequency derivative
of $\approx2\times10^{-13}$\,Hz/s, which is two orders of
magnitude larger than the frequency derivative usually observed from
millisecond pulsars of the Galactic field.  To determine the
value of the acceleration in the cluster potential, we
followed the procedure described by \citet{papitto12} using the
pulsar's radio localisation \citep{pavan13}, the measured cluster
center coordinates \citep{becker2003}, the values of the M~28 core
and tidal radii \citep{trager1995}, and its total estimated mass of
$5.5\times10^5\,\mathrm{M}_\odot$ \citep{boyles2011}.  Considering the
source spin period, and the upper limit on the period derivative
from the cluster potential, we can derive an upper limit on
the NS surface dipolar magnetic field at the equator of
$B$$\sim$3.2$\times10^{19}\sqrt{P\dot{P}} < 3.5\times10^{9}$\,G.  We
conclude that the magnetic field of \je\, is constrained to be in the
$\sim$0.7$-$35$\times10^{8}$\,G range, in line with what is predicted by
its propeller state.

Our analysis of the \xmm\ data also revealed that the aperiodic variability of \je is dominated by a flicker noise 
component extending to less than $f_{\rm min}=10^{-4}$\,Hz, similarly to black-hole binaries in the
soft states \citep[e.g.,][]{churazov2001}. This is at
odds with the low-frequency noise properties of other AMP \citep{vanstraaten2005}, 
which are usually described
by a flat topped noise with a characteristic frequency of
$\sim0.1-1$ Hz.  The hour-long time-scale of the noise observed from \je
rules out that this variability is
produced in the inner rings of the accretion disk (where most of
the X-rays come from) because the dynamical and viscous time-scales are much
shorter there.
\citet{lyubarskii1997} interpreted the flicker noise observed
from accreting compact objects in terms of variations of the
accretion rate at different radii, each characterized by a
different viscous time-scale. 
These fluctuations propagate to the inner parts of the
disk, generating the observed X-ray variability. According to
this model, the longest time-scales are introduced in the outer
parts of the disk, the radius of which is related to the
low-frequency break observed in the power spectrum by
\begin{equation}
R_\mathrm{out}\sim\left(\frac{\alpha \sqrt{GM_*}}{ f_\mathrm{min}}\right)^{2/3}
=6\times10^{10} \left(\frac{\alpha}{0.1}\right)^{2/3} \left(\frac{M}{1.4 M_*}\right)^{1/3}\,\mathrm{cm}\,,
\end{equation}
where $\alpha$ is the disk viscosity parameter of a Shakura-Sunyaev
disk. This size is smaller by a factor $\sim2$ than the size of the NS
Roche lobe:
\begin{equation}
R_{L1}=a\frac{0.49q^{-2/3}}{0.6q^{-2/3}+\log(1+q^{-1/3})}\simeq 10^{11}\,\mathrm{cm},
\end{equation}
where $a=2\times10^{11} (M_t/1.6 M_{\odot})^{1/3}$\,cm is the orbital
separation between the two stars, $M_t$ is the total mass of the
system, $q=M_2/M_1$ is the binary mass ratio, and values of $M_2=0.2$\,
M$_{\odot}$ and $M_1=1.4$\,M$_{\odot}$ were considered for the
donor and the NS mass, respectively (see paper\,I). In black-hole binaries, the RMS
variability is more pronounced at high energy ($\ga6$\,keV), suggesting
that an extended optically thin corona around the accretion
disk causes the X-ray variability. In our case, there is no significant energy dependency of 
the shape of the PDS, while coherent timing reveals a 
relatively high and constant pulsed fraction above 4\,keV. This supports the idea that 
the X-rays recorded from \je\ are produced by the accretion stream close to the NS, as 
discussed earlier in this section. 

According to our interpretation, the peculiar behavior of \je 
in the X-ray domain is mainly due to switches between accretion to weak and strong propeller states. This 
requires the magnetospheric radius to be close to the coronation radius across the entire outburst.  
A similar situation has previously been investigated in several theoretical studies
\citep{spruit93,rappaport04,dangelo10,dangelo12} and applied to interpret 
observations of a number of AMXP in outbursts. For
SAX\,J1808.4$-$3658 and NGC\,6440\,X-2 the switch between accretion and 
propeller was used to interpret periods of intense quasi-periodic flaring activity 
\citep{patruno2009,patruno2013}. Evidence that the magnetospheric and co-rotation radii 
remain locked in outburst was also reported for XTE\,J1814$-$338 \citep{haskell2011}.
\citet{patruno2012} proposed later that such locking might indeed occur during most of the AMXP outbursts.  

However, \je\ displayed a different kind of X-ray variability than that reported here. 
This was revealed during the analysis of a faint outburst from the source discovered serendipitously 
in \chandra\ archival data from 2002. During this event, \je\ reached a luminosity of $\sim$10$^{33}$~erg/s and its X-ray variability 
was ascribed to the locking of the magnetospheric radius in the proximity of the source light-cylinder
\citep{linares2014}. The same scenario was used later to interpret a period of pronounced  
low-luminosity X-ray variability observed from PSR\,J1023+0038. 
On this occasion the radio pulsations from the source 
disappeared and its GeV emission increased by a factor of $\sim$5 \citep[possibly because of outflows and the formation of shocks in the intra-binary environment;][]{takata2014,patruno2014}. 
A similar behavior was also observed from the source XSS~J12270$-$289 \citep{papitto2014,bassa2014,roy2014}.
Millisecond pulsars in binary systems thus seem to display frequent episodes of mass accretion during which 
the magnetospheric boundary can either be locked at the light cylinder or at the co-rotation radius. In the former case, 
accretion is strongly inhibited and the system is observed as a relatively faint X-ray emitter ($L_X\sim10^{32-33}$\,erg/s); 
in the latter, case canonical outbursts might occur and peak X-ray luminosities of $L_X\sim10^{35-36}$\,erg/s can be achieved.

\section{Conclusions}
\label{sec:conclusions}

We analyzed the peculiar variability showed in the X-ray domain by the swinging pulsar \je.\ 
We discussed that the results obtained from the spectroscopic and the timing analysis of the \xmm\ data could be reasonably well described by assuming 
an unstable accretion phase in which the material from the disk does not always reach the NS surface quietly, but proceeds in a ``hiccup'' fashion: 
accretion is strongly inhibited for a substantial fraction of time by the stellar rotation that causes the onset of a propeller regime. 
This might produce strong outflows that generate a strongly variable GHz emission and lead to the sporadic appearance of a completely different X-ray spectrum than that of
the remaining accretion phases. 

Observations of possible future outbursts from \je\  
with the next generation of X-ray timing experiments \citep[e.g, LOFT;][]{feroci12} will enable us 
to perform timing and spectral analyses with higher statistics down to very short time-scales,  
providing crucial insights into the mechanisms responsible for the peculiar X-ray variability of this source and other AMXPs.  

\section*{Acknowledgements}
The authors thank the \xmm, \swift, and ATCA PIs and the respective shift teams for their availability 
in quickly performing  the target of opportunity observations used in this work. 
This work made use of data supplied by the UK Swift Science Data Centre at the 
University of Leicester. AP and NR acknowledge grants AYA2012-39303 and SGR2009- 811. AP is
supported by a Juan de la Cierva fellowship. NR is supported by a
Ramon y Cajal fellowship and by an NWO Vidi Award.
We thank our colleagues M. Audard, M. Guainazzi, E. Kuulkers, S. E. Motta, A. Patruno, and A. Tramacere, 
for contributing with stimulating discussions and advices.

\bibliographystyle{aa}
\bibliography{J18bib}

\begin{appendix}
\section{analysis of the full-resolution \rgs spectra}
\label{sec:appendix}

During the analysis of the full-resolution \rgs spectra of \je\ in obs2 (which have a much longer exposure time than
obs1, see Table.\,\ref{tab:obs}), we noticed a convincing hint of emission lines from \ion{O}{VII} around 22\,\AA. 
Because these lines do not affect our discussion and conclusion about the source spectral and timing variability reported 
in Sect.\,\ref{sec:discussion} and \ref{sec:conclusions}, we summarize this result here. 

\begin{figure}
  \begin{center}
        \includegraphics[angle=270,width=0.45\textwidth]{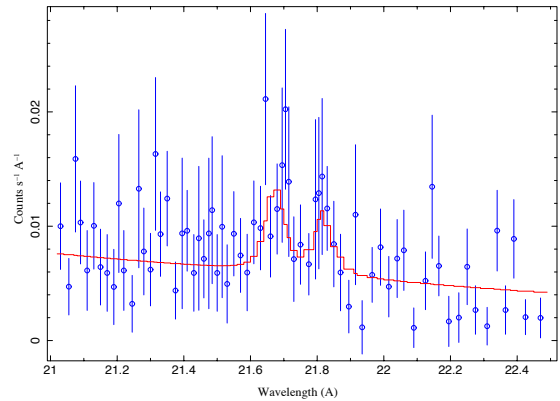}
        \caption{Zoom of the \rgs2 spectrum of obs2 rebinned to achieve an S/N of 3 in each bin around the \ion{O}{VII} triplet. The width of lines is fixed to zero.}
\label{fig:spectrum_lines}
\end{center}
\end{figure}

Fig.\,\ref{fig:spectrum_lines} shows the \rgs2 spectrum in obs2 after rebinning 
to achieve a minimum S/N of 3.0 in each wavelength bin)\footnote{We used the ISIS software package (version 1.6.2-16) 
for the spectral analysis presented in this section.}. 
Table~\ref{tab:lines} gives the best determined parameters obtained by fitting the 
two \rgs spectra in obs2 with a model comprising the continuum determined in the broad-band analysis (Sect.~\ref{sec:spectral}) 
plus two narrow Gaussian lines 
\citep[we used the $\chi^2$ statistics with weights according to][]{gehrels1986}. 
The wavelengths of the two emission lines were well constrained, while only an upper limit 
was obtained on their width. No forbidden line (22.10\,\AA) or lines typical of other isotopes 
(\ion{C}{V}, \ion{N}{VI}, \ion{Ne}{X}, \ion{Mg}{XI}, and \ion{Si}{XII}) were significantly detected. 

The two emission lines correspond to the oxygen triplet. The intercombination line is at the expected wavelength 
(the upper limit on its Doppler broadening is about 700\,km/s), while  the resonance line
appears to be slightly redshifted and the upper limit on its width is larger. 
The absence of a forbidden line can either be due to the high density of the gas ($N_e\ga10^{11}\,\mathrm{cm}^{-3}$)
or to the presence of an intense UV field in the line forming region \citep{porquet2000,porquet2001}, as expected in the proximity of the X-ray irradiated companion star \citep{pallanca13}. 
If we assume that the lines can be broadened by the plasma orbital motion, 
the upper limit on the width of the intercombination line is compatible with the Keplerian velocity at the orbital separation 
of the system, while the slightly broader resonance line would favor emission from an inner region of the disk. 
This constraint is not stringent, however, as it was suggested that emission lines that originate in the outer layers 
of an accretion disk might be narrower than naively expected \citep{garate2002,garate2005}.
We remark that the relatively low statistics of the \rgs spectrum do not allow us to perform an orbital phase-resolved analysis
of the oxygen lines. 
\begin{table}
\caption{Best-fit model parameters of \ion{O}{VII} lines during obs2.}
\begin{center}
\begin{tabular}{cr@{}lc}
\hline
\hline
\smallskip
$N_\mathrm{nthComp}$\tablefootmark{a}  & 0.032 \\
\smallskip
$\lambda_\mathrm{I}$ & 21.82 &$^{+0.02}_{-0.03}$ & \AA  \\
\smallskip
$\sigma_\mathrm{I}$ & $<$&0.05\tablefootmark{b} & \AA  \\
\smallskip
$A_\mathrm{I}$ & 10 &${_-7}^{+8}$ & $10^{-5}\mathrm{ph\,s^{-1}\,cm^{-2}}$  \\
\smallskip
$\lambda_\mathrm{R}$ & 21.675 &$^{+0.034}_{-0.042}$ & \AA  \\
\smallskip
$\sigma_\mathrm{R}$ & $<$&0.09 & \AA  \\
\smallskip
$A_\mathrm{R}$ & 13 &$\pm7$ & $10^{-5}\mathrm{ph\,s^{-1}\,cm^{-2}}$  \\
\smallskip
$\lambda_\mathrm{F}$ & 22.10 &- & \AA  \\
\smallskip
$\sigma_\mathrm{F}$ & 0&- & \AA  \\
\smallskip
$A_\mathrm{F}$ & $<$ & 1.8 & $10^{-5}\mathrm{ph\,s^{-1}\,cm^{-2}}$  \\
\smallskip
$\chi^2_\mathrm{red}$/d.o.f & 1.01 & /64 & \\
\hline
\end{tabular}.
\tablefoot{The subscripts $I$, $R$, and $F$ refer to the intercombination, resonance, and forbidden lines, respectively. The fit has been performed fixing the line width 
to zero, while the upper limit was determined in a second iteration with the line width set as a free parameter.\\
\tablefoottext{a}{The normalization of the continuum is first recomputed in the fit and then fixed; all the other continuum parameters are fixed to those of 
the average spectrum measured in Sect.\,\ref{sec:spectral}.}}
\end{center}
\label{tab:lines}
\end{table}

\section{analysis of the ATCA observation of \je\ }
\label{sec:appendix2}

In this appendix, we report on the analysis of the radio observation performed in the 
direction of \je\ by the Australia Telescope Compact Array (ATCA; Target of Opportunity 
project CX258) for $\sim$6 hours on 2013 April 5. 
The preliminary results of this observation were reported by \citet{pavan13} and are summarized in 
paper\,I, but the data analysis was not presented in full detail yet. 
 
The data considered here were collected with the new Compact Array
Broadband Backend \citep[CABB,][]{wilson2011} in the hybrid array
configuration H214 at wavelengths of 6 and 3~cm ($\nu$=5.5 and 9~GHz). 
The observation was carried out in dual-frequency mode, with $\sim$3.7 hours of integration time,
split into two periods separated by approximately 2.5 hours.
The primary calibrator was \object{PKS~B1934$-$638}, while
\object{PKS~1817$-$254} was used for phase calibration. The data were
analysed with the \textsc{miriad} \citep{sault1995} and \textsc{karma} \citep{karma1996} 
software packages distributed by ATNF. 

Radio images of the observation were created using the \textsc{miriad} multifrequency
synthesis and robust weighting \citep{sault1995}. They were deconvolved
using the {\sc mfclean} and {\sc restor} algorithms with primary-beam
correction applied using the {\sc linmos} task. A similar procedure
was used for the \textit{U} and \textit{Q} Stokes parameter
maps. 

A single compact source was detected in our images within the M28 globular cluster 
(the image resolution was 4.8\arcsec$\times$1.0\arcsec\ at PA=9 deg with an
estimated r.m.s. noise of 11\,$\mu$Jy/beam at 5.5GHz and 3.0\arcsec$\times$0.64\arcsec\ at PA=10 $\deg$; with an
estimated r.m.s. noise of 12\,$\mu$Jy/beam at 9~GHz). Its position coincides well with the X-ray counterpart of PSR J18245-2452 (Paper I).  
The measured flux density of the source was $(0.62\pm0.03)$\,mJy at 5.5~GHz and $(0.75\pm0.04)$\,mJy at 9~GHz \citep{pavan13}.
The radio emission recorded from \je\ therefore has an intensity similar to that of other AMXPs in outburst \citep{migliari2006,patruno12}.  

\begin{figure}
\begin{center}
\includegraphics[width=0.5\textwidth]{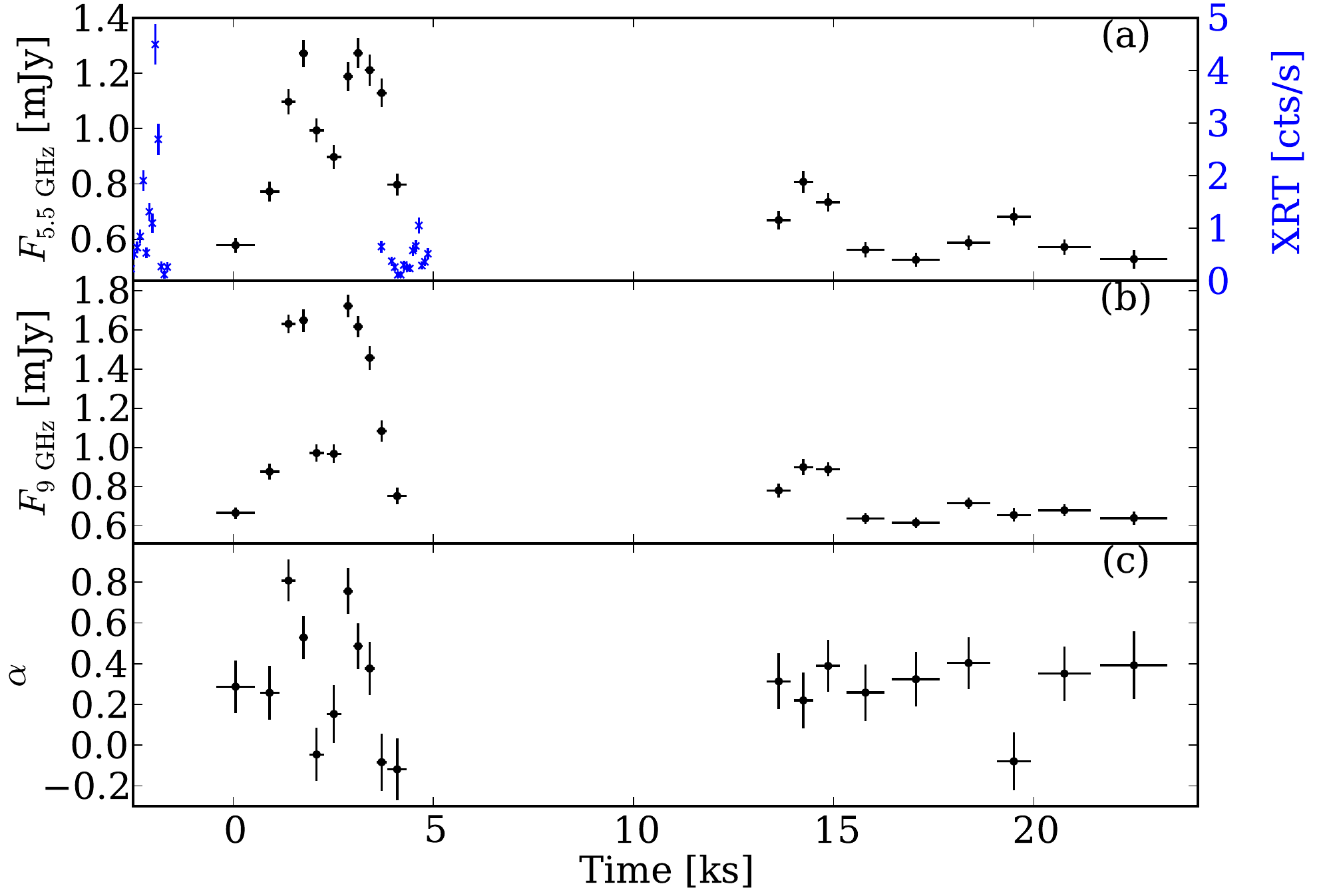}
\caption{ATCA lightcurve of \je\ in the two frequencies (5.5\,GHz, panel (a), and 9\,GHz, panel (b)) and the calculated spectral index $\alpha$, panel (c). 
We also reported in the upper panel the 0.5--10\,keV lightcurve  from the \swift\,/XRT observation ID.\,00032785003 (in blue). 
These were the only X-ray data collected quasi-simultaneously with the ATCA observation.}
\label{fig:atca}
\end{center}
\end{figure}

We extracted radio light curves of the source at the two observing frequencies by using the \textsc{miriad/uvplt} task, and
averaging the data at temporal intervals of 2~minutes. The source showed variations of intensity during the first 
part of the observation, reaching up to 2~mJy, while during the second part, the mean flux density was not significantly variable (Fig.~\ref{fig:atca}).
To better study the observed radio variability, we estimated
the variation of source spectral index $\alpha$ with time. The lightcurves and
spectral index are shown in Fig.\,\ref{fig:atca}. The spectral index varies significantly in the range $-0.2$--0.8 and has a
mean value of $\alpha\sim+0.39\pm0.21$ (where $\alpha$=$\left[\log(F_{\nu1}/F_{\nu2})\right]/\left[\log(\nu1/\nu2)\right]$, $\nu1$=9\,GHz, and $\nu2$=5.5\,GHz).
Splitting the two observational bands into bins of bandwidth 128~MHz
allowed us to obtain a refined mean value  of the spectral index $\alpha = +0.39 \pm 0.06 $. 
This inverted radio spectrum indicates the \je\ has an optically thick synchrotron-emitting 
region or a mixture of thermal and synchrotron emission, typical of a persistent jet or a less collimated 
outflow \citep[e.g.,][]{seaquist1993}.

We report in Fig.~\ref{fig:atca} the only available quasi-simultaneous X-ray observation of the source. 
These data (observation ID.\,00032785003) were collected with the \swift\,/XRT on 2013 April 5 from 16:58:06 to 19:04:26 (UTC) and  
analyzed with the Leicester XRT on-line tool\footnote{See \url{http://www.swift.ac.uk/user_objects}.} \citep{evans09}.  
The observation is divided into two snapshots that occur before the start of the radio data and at the end of the second radio flare.
The spectra of the two snapshots were extracted separately and then fit together with an absorbed power-law model.
The first snapshot was affected by pile-up and thus the corresponding spectrum was extracted by using an annular region with inner 
radius of 4\,pixels and external radius of
20 pixels, in agreement with standard procedures \citep{burrows05, romano2006}.
The photon indices of the power laws were left free to vary in the fits, while the absorption column density
was constrained to be the same for the two spectra.
The best-fit model
($\chi^2_\mathrm{red}=0.9$, 32 d.o.f.)
gave an absorption column density of
 $N_\mathrm{H}=(4.1\pm1.6)\times 10^{21}\,\mathrm{cm}^{-2}$ and
photon indices of
$(1.45\pm0.25)$ and $(1.8\pm0.3)$,
for the two snapshots.
The corresponding fluxes (0.5-10\,keV) were
$(6.4\pm0.8)$ and $(4.3\pm0.6) \times 10^{-11}\,\mathrm{erg/s/cm^2}$.
We note that the spectral properties and fluxes of the two XRT snapshots performed simultaneously with the ATCA observation 
are similar to those measured during the soft magenta states identified by XMM (see Table~5). 
The slightly steeper power-law index of the second snapshot might be generated by
black-body emission, which went undetected owing to the limited number of collected photons.
These results support the idea that outflows might be produced during the magenta states of \je.\ 

\end{appendix}

\end{document}